\documentclass[final,3p,authoryear,11pt]{elsarticle}

\usepackage{fancyhdr}
\usepackage{amsmath,amssymb}
\usepackage{graphicx}
\usepackage{booktabs}
\usepackage{multirow}
\usepackage{hyperref}
\usepackage{caption}
\usepackage{float}
\usepackage{subcaption}
\usepackage{geometry}
\usepackage{threeparttable}
\usepackage{setspace}
\usepackage{float}
\usepackage{longtable}
\usepackage[ruled,vlined,linesnumbered]{algorithm2e}

\usepackage{amsthm}
\usepackage{booktabs}   
\usepackage{tabularx}   
\usepackage{array}      
\usepackage{ragged2e}   
\usepackage{makecell}   
\newtheorem{assum}{Assumption}

\newtheorem{constraint}{Constraint}
\newcolumntype{L}[1]{>{\RaggedRight\arraybackslash}p{#1}}
\newcolumntype{Y}{>{\RaggedRight\arraybackslash}X}

\hypersetup{
    colorlinks=true,
    linkcolor=blue,
    citecolor=blue,
    urlcolor=blue
}

\journal{Transportation Research Part A: Policy and Practice}

\begin{document}

\begin{frontmatter}

\title{Stochastic Dynamic Pricing of Electric Vehicle Charging with Heterogeneous User Behavior: A Stackelberg Game Framework}

\author[1]{Yongqi Zhang}
\author[1]{Dong Ngoduy\corref{cor1}}
\cortext[cor1]{Corresponding author: Dong Ngoduy, Email: Dong.Ngoduy@monash.edu}
\author[2]{Li Duan}
\author[1,3]{Mingchang Zhu}
\author[1]{Zhuo Chen}

\address[1]{Department of Civil and Environmental Engineering, Monash University, Melbourne, Australia}
\address[2]{School of Civil and Hydraulic Engineering, Huazhong University of Science and Technology, Wuhan, China}
\address[3]{The Key Laboratory of Road and Traffic Engineering, Ministry of Education, Tongji University, Shanghai, China}

\begin{abstract}
The rapid growth of electric vehicles (EVs) introduces complex spatiotemporal demand management challenges for charging station operators (CSOs), exacerbated by demand imbalances, behavioral heterogeneity, and system uncertainty. Traditional dynamic pricing models, which rely on deterministic EV-CS pairings and network equilibrium assumptions, often oversimplify user behavior and lack scalability. This study proposes a stochastic, behaviorally heterogeneous dynamic pricing framework formulated as a bi-level Stackelberg game. The upper level optimizes time-varying pricing to maximize system-wide utility, while the lower level models decentralized EV users via a multinomial logit (MNL) choice model that incorporates price sensitivity, battery aging, risk attitudes, and network travel costs, explicitly avoiding network equilibrium constraints to enhance scalability. Congestion effects are represented using queuing-theoretic approximations for waiting times and rejection probabilities. To efficiently solve the resulting large-scale, time-varying optimization problem while preserving the leader-follower structure, a rolling-horizon solution approach combining Dynamic Probabilistic Sensitivity Analysis-guided Cross-Entropy Method (PSA-CEM) with the Method of Successive Averages (MSA) is implemented. A real-world case study in Clayton, Melbourne, with 22 charging stations (19 fast-charging and 3 Level 2), validates the framework. Simulation results demonstrate that the proposed dynamic pricing mechanism substantially reduces queuing penalties, improves EV user utility, and increases overall system performance compared with fixed and time-of-use pricing. The framework adapts to station heterogeneity and demand fluctuations, providing a robust and scalable tool for strategic and operational EV charging management. By integrating dynamic pricing, stochastic user choice, and congestion-aware modeling, the approach enables adaptive, large-scale EV charging optimization while balancing realism with computational efficiency.
\end{abstract}


\end{frontmatter}

\section{Introduction}
Electric vehicles (EVs) have significant advantages over traditional vehicles in terms of environmental friendliness, energy efficiency, and low maintenance requirements. In recent years, with the growing emphasis on sustainable development, many countries have actively promoted EV adoption through subsidies and tax incentives. Meanwhile, the layout of charging stations (CSs) has been continuously improved in both urban and rural areas, and EVs and their charging infrastructures have developed rapidly. By the end of 2024, the total number of EVs on roads worldwide had exceeded 58 million, with approximately 5.45 million public chargers installed globally \citep{IEA2025GlobalEV}. The rapid surge in EV charging demand has brought considerable challenges to the planning and operation of charging systems. However, this expansion trend also introduces new challenges. 

Firstly, the continuous growth of EV ownership has outpaced the construction speed of new CSs \citep{afirev2022observatoire}. Secondly, a rational spatial configuration of charging stations is crucial for the sustainable development of the EV ecosystem, yet many existing CS layouts remain suboptimal or unscientifically distributed \citep{feng2023evfcsplanning}. More importantly, current charging pricing strategies for EVs require more intelligent and adaptive mechanisms that can balance the profitability of charging station operators (CSOs) with the welfare of EV users, thereby ensuring a reasonable regulation of charging demand \citep{cui2023drlpricing}. The location, capacity, and pricing strategy of CSs significantly influence EV users' charging choices, which in turn affect CSOs' pricing decisions. This interaction often leads to supply-demand imbalances and reduced system efficiency under demand uncertainty \citep{Kazemtarghi2024,BenGharbia2023,Patel2022,WANG2025104666,Amilia2022}. For instance, some CSs may experience severe congestion with long queues of waiting EVs, while others remain underutilized, resulting in grid load imbalance and reduced profitability for CSOs. Therefore, an effective charging pricing mechanism is essential to guide EVs toward less congested stations and to achieve optimal resource allocation.

Establishing a demand-driven dynamic pricing mechanism has thus become an urgent necessity \citep{Ali2023}. Existing Stackelberg game-based models consider factors such as price elasticity and travel distance, yet they often overlook the influence of actual traffic routes \citep{Lee2019}. Other studies employ multi-agent reinforcement learning or stochastic optimization frameworks, but they either lack explicit, closed-form EV-CS attraction models or fail to adequately capture queuing congestion effects \citep{Lu2022}. In the transportation research domain, scholars frequently adopt bi-level optimization models to couple transportation and power systems \citep{Zhou2021,Tran2021b,Song2025}. In such frameworks, the upper level aims to maximize overall system welfare, while the lower level seeks user equilibrium or system-optimal traffic assignment. Although these models provide a detailed understanding of drivers' route choices and network congestion, they remain computationally intensive and challenging to solve for large-scale systems \citep{Mi2023,Yu2021}.

Unlike road congestion, which arises from interactions among all vehicles across the urban traffic network, congestion at charging stations mainly manifests as excessive waiting time in the charging service process. Such queuing primarily affects the quality of charging services and, consequently, the experienced utility of EV users, while also influencing station-level operational efficiency \citep{grigorev2021evimpact,Tran2021a}. 

Moreover, charging station queuing is largely driven by the charging demand of EVs and the service capacity of charging infrastructure, rather than by citywide traffic conditions. Therefore, this study focuses on optimizing the EV-CS matching mechanism and dynamic pricing strategy at the urban scale, without explicitly modeling general traffic congestion.

To tackle these challenges in a tractable manner for large-scale urban networks, we adopt a bi-level optimization framework as in \citep{Zhou2021,Tran2021b,Song2025}. More specifically, in our framework, the upper level aims to maximize overall system utility by considering both CS and EV benefits, while the lower level minimizes EV losses from queuing overflow under stochastic and heterogeneous travel conditions.

To reduce computational complexity and facilitate large-scale analysis, the lower-level traffic network is modeled using a simplified charging demand and service queuing representation, which captures the essential interactions between EV arrivals and station capacity without explicitly simulating citywide traffic flows. This approximation enables efficient evaluation of the lower-level outcomes while preserving the accuracy of the upper-level optimization.

In summary, our scientific contributions to the literature are given below.
\begin{enumerate}
    \item We provide a comprehensive characterization of EV random utility arising from imperfect information and perception biases in charging station selection, while explicitly accounting for the heterogeneity of both EVs and CSs.
    \item We develop a Stackelberg game-based dynamic pricing framework in which the upper-level government authority determines time-varying prices to maximize system-wide utility, and the lower-level EV users make probabilistic charging station choices under stochastic and heterogeneous travel and congestion conditions, forming a stochastic choice equilibrium while considering capacity constraints and potential station overloading.
    \item We propose a rolling-horizon heuristic algorithm to capture the stochastic and time-varying nature of the decision-making process. Numerical experiments demonstrate that the approach efficiently identifies high-quality solutions, providing a practical tool for planners to incorporate realistic operational considerations.
\end{enumerate}

The paper is structured as follows: Section II reviews the bi-level optimization problem used for the dynamic pricing strategy and identifies research gaps. Section III presents our methodology and the solution algorithms for the proposed optimization problem. Numerical experiments are provided in Section IV.

\section{Literature Review}
\subsection{EV-CS Dynamic Pricing Problem}

Electric vehicles, as a critical component of carbon-reduction strategies, have garnered extensive attention from both academia and industry~\citep{Song2025}. With the rapid growth in EV market penetration, optimizing charging pricing strategies to guide user behavior has emerged as a central research topic~\citep{Gupta2023}. However, the surge in EV ownership has introduced significant challenges, particularly uncoordinated charging behavior that causes congestion at charging stations and grid overloading~\citep{Bayram2015,Chen2024Etiquette}. Dynamic pricing, which adjusts charging prices in response to real-time supply-demand conditions, has been widely recognized as an effective mechanism to mitigate these issues~\citep{Saharan2020}.

Existing research on EV charging dynamic pricing can be broadly categorized into three modeling paradigms: centralized optimization, distributed or coordinated optimization, and hybrid approaches represented by game-theoretic models~\citep{Amjad2018}.

Centralized methods formulate a single optimization problem for the entire system, enabling globally optimal solutions but often becoming computationally intractable for large-scale networks~\citep{Wang2021,Amjad2018}. Distributed methods, in contrast, decompose the problem across multiple agents (e.g., individual stations or EVs), allowing each agent to make local decisions while coordinating with others to achieve system-level objectives; these frameworks are more scalable and preserve privacy but require effective coordination~\citep{Qin2011,Amjad2018,Popiolek2023}. Hybrid approaches, often based on game-theoretic models such as Stackelberg games, combine centralized control with distributed decision-making: leaders (e.g., charging operators) set prices or policies, and followers (e.g., EV users or stations) respond optimally~\citep{Sheng2021,Hu2023}. Representative studies include multi-leader Stackelberg games modeling distinct objectives of traffic and power operators~\citep{Sheng2021}, three-stage frameworks analyzing the impact of private charging infrastructure on public markets~\citep{Hu2021}, and three-level game structures involving governments, platforms, and private owners for subsidy design~\citep{Wang2025}.

Despite their success, Stackelberg-based bilevel models continue to face significant modeling and computational challenges. The upper-level problem is often nonconvex and nondifferentiable due to its implicit dependence on lower-level equilibrium, while repeated computation of user equilibrium in each iteration leads to exponential growth in computational cost, limiting scalability to large-scale applications. Moreover, most studies rely on deterministic assumptions—assuming perfectly rational users and known charging demand—whereas real-world behavior exhibits stochasticity and heterogeneity driven by incomplete information, perception bias, and demand uncertainty. To address these issues, various computational approaches have been explored. Mathematical programming has been widely adopted for both single- and multi-stage pricing problems, though methods such as stochastic optimization often neglect critical factors like queuing delay, which is essential for capturing feedback between congestion and user choice~\citep{Kazemtarghi2024}. Receding-horizon control methods, such as model predictive control (MPC), extend these models to online applications~\citep{Zheng2019} but typically optimize single expected scenarios, making it difficult to explicitly capture stochastic variability and uncertainty propagation. For large-scale or highly nonlinear problems, heuristic algorithms (e.g., genetic algorithms, particle swarm optimization) have been employed to approximate Stackelberg equilibria~\citep{Tran2021b,Tran2023,Yin2024}, though they typically yield locally optimal solutions without convergence guarantees. More recently, deep reinforcement learning (DRL) has emerged as a data-driven alternative capable of deriving adaptive Stackelberg strategies without explicit analytical formulations~\citep{Lu2022}, though they often lack closed-form explanations of how EVs are allocated among stations.

In summary, while substantial progress has been achieved, existing studies still suffer from several critical limitations, including high computational complexity, lack of global optimality guarantees, and insufficient representation of uncertainty and user behavioral heterogeneity.

\subsection{Stochasticity Levels in EV Charging and Pricing Models}

Stochasticity arises in EV charging and pricing systems at multiple levels, reflecting the complex and uncertain nature of user behaviors, external environments, and system dynamics. From a modeling perspective, existing studies can be broadly categorized into three representative levels of stochastic representation: behavioral, scenario-based (external), and process-level stochasticity.

At the microscopic level, behavioral stochasticity captures users' bounded rationality and heterogeneous perceptions. It is typically modeled through discrete choice formulations such as the Logit model with perception errors~\citep{Soares2017Dynamic}, which describe imperfect information and probabilistic decision-making in EV users' selection of charging stations.

At the mesoscopic level, scenario-based stochasticity (also referred to as external environment stochasticity) represents exogenous uncertainties—such as renewable generation, electricity prices, and EV charging demand—using a finite set of probabilistic scenarios. Representative studies optimize dynamic pricing under uncertain conditions of battery electric vehicle (BEV) demand, photovoltaic (PV) output, and frequency-regulation signals to achieve robust day-ahead operation~\citep{Wu2019Stochastic}, or incorporate predefined uncertainties in EV arrival rates, charging power, and energy demand~\citep{Kazemtarghi2024}. By optimizing over multiple probabilistic realizations of the system, these models achieve expected or risk-averse optimality, ensuring that pricing strategies remain near-optimal across varying environmental and market conditions~\citep{Elkholy2026}.

At the macroscopic level, process stochasticity characterizes the temporal evolution of system states driven by random process noise, typically within stochastic dynamic programming (SDP) or Markov decision process (MDP) frameworks. Unlike scenario-based models, where uncertainty is predefined and static, process-level stochasticity emphasizes sequential decision-making under evolving uncertainty~\citep{Luo2018Stochastic, Zanvettor2022Stochastic}. The randomness is gradually revealed over time, allowing adaptive control of dynamic pricing, energy storage, and grid interactions in response to real-time disturbances. Such frameworks effectively extend scenario-based models by embedding temporal coupling and recursive optimization structures (e.g., Bellman recursion), providing a more realistic depiction of dynamic market and system responses.
Overall, these three layers of stochastic representation collectively describe the intrinsic uncertainty structure of dynamic pricing systems—from individual user choice randomness to external environmental fluctuations and temporally evolving process dynamics. Integrating these dimensions offers a comprehensive understanding of how stochastic modeling supports robustness, adaptivity, and realism in EV charging and pricing optimization.

\subsection{Heterogeneity Levels in EV Charging and Pricing Models}

In the modeling and optimization of electric transportation systems, heterogeneity manifests across multiple dimensions, reflecting the diversity of participants and the complexity of their interactions in real-world systems.

First, at the vehicle level, researchers typically consider differences among various types of electric vehicles in terms of energy efficiency, driving range, and battery capacity. These factors directly affect routing and charging demands, thereby influencing energy distribution and travel equilibrium within the system~\citep{Mrkos2022,Elkholy2026}.

Second, at the infrastructure level, scholars have investigated the heterogeneity of charging station types. For example, chargers of different power levels (standard, fast, and ultra-fast) exhibit significant disparities in investment costs and service efficiency, which affect station layout and optimal investment decisions. Studies have shown that introducing multiple charger types can improve model performance; however, the high cost of fast chargers tends to limit the total number of stations that can be installed, and the addition of ultra-fast chargers does not necessarily yield significant improvements in system efficiency~\citep{Park2024,Kullman2021}.

Third, at the operator level, some studies have jointly considered heterogeneous fleets and non-identical charging facilities. For instance, \citet{Calik2021} addressed the electric vehicle location-routing problem (EV-LRP) by simultaneously modeling differences in vehicle types, charging facilities, and partial recharging behavior. They proposed a two-phase Benders decomposition algorithm that efficiently solves the problem under realistic operational settings, providing both methodological and managerial insights into heterogeneous EV logistics systems.

Fourth, at the user level, heterogeneity manifests in differences in income and value of time (VOT). \citet{Zheng2020} incorporated income- and VOT-based heterogeneity into a congestion pricing framework to explicitly capture differentiated responses of user groups to tolling. By modeling populations with Gini-index-based VOT distributions, they demonstrated that VOT-based differentiated pricing can substantially enhance equity among heterogeneous travelers while maintaining near-system-optimal efficiency.

Finally, at the multi-agent interaction level, recent studies have extended heterogeneity to complex intelligent systems. \citet{Mao2024} proposed a heterogeneous multi-agent hypergraph attention actor-critic (HMA-HGAAC) framework that captures inherent differences between EVs and battery swapping stations (BSSs) in terms of state space, objectives, and decision logic. This enables each agent type to make independent yet coordinated decisions within a unified learning framework, significantly improving system cooperation and global optimality. Similarly, \citet{Zhang2022} modeled the heterogeneity among CSOs in terms of geographic distribution, user demand, pricing strategy, and energy access, achieving more personalized and fine-grained dynamic pricing and scheduling strategies.

In summary, heterogeneity has become a key modeling element in the optimization of electric transportation systems. Its multi-level manifestations—across vehicles, facilities, fleets, users, and intelligent agents—not only enhance the realism and representational power of models but also lay the foundation for future research in fairness, efficiency, and coordinated optimization. However, it is worth noting that in dynamic electric charging pricing models, relatively few studies simultaneously incorporate heterogeneity across vehicles and facilities, particularly under the joint consideration of behavioral stochastic model components.

\subsection{Research Gaps}

Although the dynamic pricing problem of EVs and its extensions on stochasticity and heterogeneity modeling have been extensively investigated, several critical research gaps remain from both modeling and methodological perspectives.

First, at the dynamic charging pricing modeling level, although the Stackelberg game has become a mainstream framework for EV-CS pairing and dynamic pricing problems, existing models often rely on simplifying assumptions. Most studies treat the upper level as a single charging operator or regulator, while the lower level focuses solely on rational equilibrium responses of EV users, neglecting cross-layer feedback between price signals, spatiotemporal travel patterns, and queuing dynamics. Furthermore, to maintain computational tractability, many models adopt deterministic or static formulations, which fail to capture path dependency and temporal coupling inherent in dynamic pricing adjustments. From a computational viewpoint, traditional bi-level programming and heuristic algorithms can obtain feasible solutions for small- or medium-scale problems but suffer from heavy computational burdens and convergence instability in large-scale, multi-agent, or multi-stage systems. Although the Cross-Entropy Method (CEM) has shown significant advantages in tackling complex, stochastic, and multi-stage optimization problems, particularly in large-scale combinatorial settings due to its systematic exploration of the decision space, it can be computationally expensive as problem size grows. Therefore, developing generalizable, more efficient, and interpretable solution approaches remains an open research challenge.

Second, at the stochastic modeling level, prior studies have predominantly emphasized exogenous scenario-based uncertainties (e.g., renewable generation, electricity prices, or charging demand) or process-level stochastic evolution. However, less attention has been paid to behavioral stochasticity arising from EV users' bounded rationality and perception bias. As the lower-level model in this study focuses on individual EV decisions in charging-station selection, behavioral stochasticity is considered the primary stochastic dimension. Unlike previous deterministic user equilibrium assumptions, this perspective explicitly recognizes users' probabilistic and imperfect decision-making under uncertain price and congestion conditions, providing a more realistic behavioral foundation for upper-level dynamic pricing optimization.

Finally,  at the heterogeneity modeling level, although existing studies have introduced heterogeneity across vehicles, facilities, users, and agents, there is still a lack of models that jointly incorporate vehicle and facility heterogeneity under stochastic dynamic pricing frameworks. In most cases, heterogeneity and stochasticity are treated separately: heterogeneous models assume deterministic behavior, while stochastic models often rely on homogeneous agents. This separation limits the representational capacity and behavioral realism of dynamic pricing frameworks in real-world electric mobility systems. Hence, future research should aim to establish an integrated paradigm that simultaneously captures multi-source heterogeneity and behavioral stochasticity within a hierarchical structure, enabling more robust, fair, and adaptive pricing decisions.

In light of these limitations, this study proposes a novel stochastic and heterogeneous Stackelberg dynamic pricing framework for the EV-CS system. The framework explicitly models behavioral stochasticity at the lower level to capture bounded rationality and perception heterogeneity in EV charging choices, while integrating vehicle- and facility-level heterogeneity into upper-level pricing design. This unified representation not only enhances the interpretability of user responses and market dynamics but also contributes to the development of more efficient, equitable, and low-carbon electric mobility systems.

\section{Methodology}

\begin{figure}[t]
    \centering
    \includegraphics[width=\linewidth]{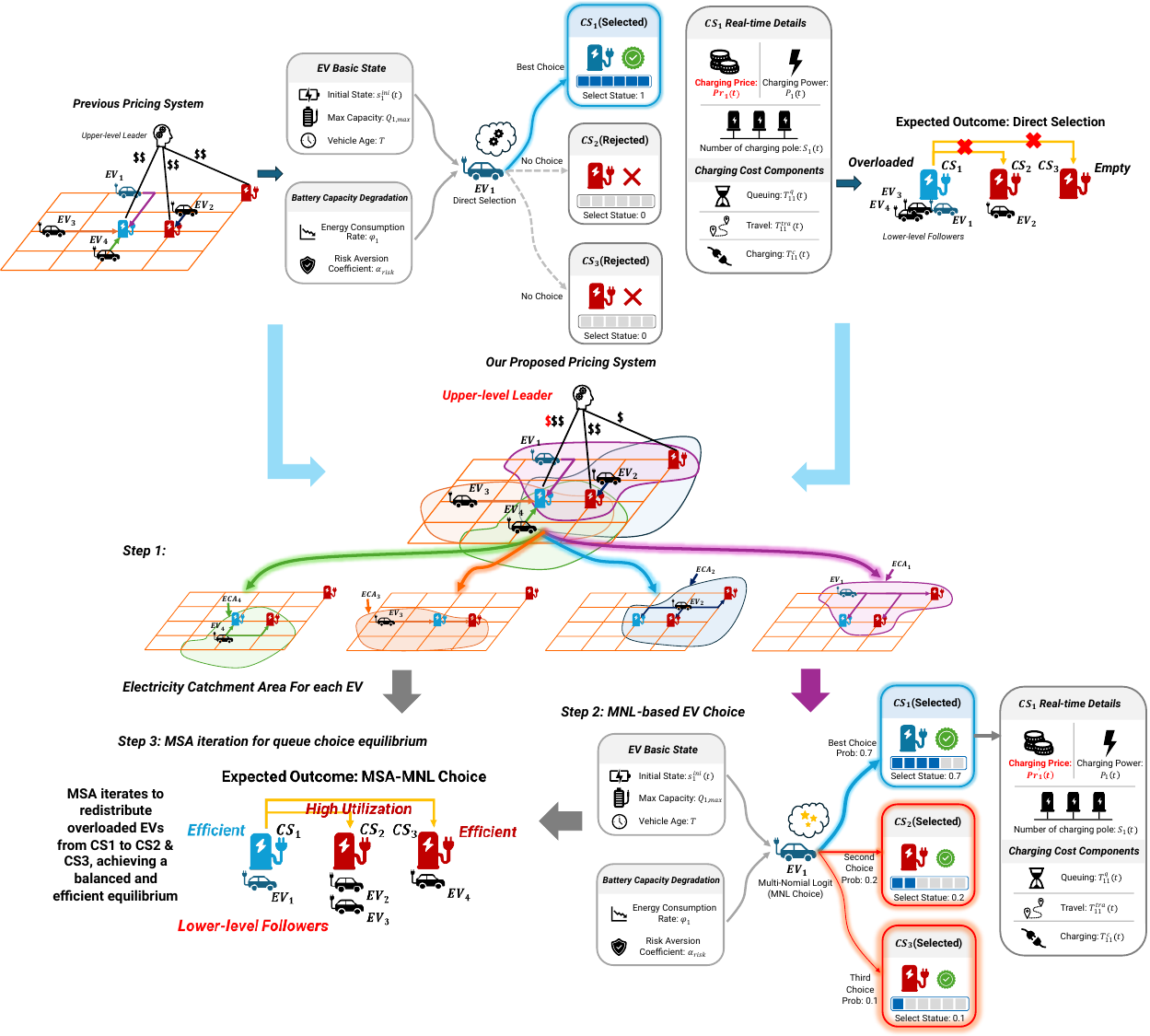}
    \caption{Overall framework of the proposed bi-level Stackelberg optimization for stochastic EV charging station selection and dynamic pricing.}
    \label{fig:framework}
\end{figure}

This section introduces the system model and the associated solution approach, comprising three key components, as illustrated in Figure~\ref{fig:framework}. 
First, a novel stochastic dynamic user choice equilibrium model is proposed, which integrates a charging station attractiveness model with queuing theory to capture EV users' rational charging behavior and the competition among charging stations. 
Second, a bi-level Stackelberg game framework, together with its associated constraints, is formulated in a sequential manner. 
Third, a dynamic pricing optimization model with a rolling-horizon scheme is developed to solve the Stackelberg game, explicitly accounting for user heterogeneity and demand stochasticity. 
The notation used in the ensuing sections is summarized in Table~\ref{tab:notation}.

\vspace{-1em}
\begin{longtable}{lp{0.8\linewidth}}
\caption{Summary of Notation}\label{tab:notation} \\
\hline
\textbf{Symbol} & \textbf{Explanation} \\
\hline
\endfirsthead

\multicolumn{2}{c}%
{{\tablename\ \thetable{} -- continued from previous page}} \\
\hline
\textbf{Symbol} & \textbf{Explanation} \\
\hline
\endhead

\hline \multicolumn{2}{r}{{Continued on next page}} \\
\endfoot

\hline
\endlastfoot

\multicolumn{2}{l}{\textit{Sets and Indices}} \\
$i$ & Index of charging station (CS) \\
$j$ & Index of electric vehicle (EV) \\
$t$ & Time index \\
$N_s$ & Total number of charging stations \\
$N_h$ & Total number of time periods (hours) \\
$\mathrm{ECA}_j(t)$ & Catchment area (reachable CSs) for EV $j$ at time $t$ \\
$\widetilde{\mathrm{ECA}}$ & Adjusted ECA considering risk and battery aging \\

\multicolumn{2}{l}{\textit{EV and Charging Parameters}} \\
$Attr_{ij}(t)$ & Attractiveness of CS $i$ to EV $j$ at time $t$ \\
$p_{ij}(t)$ & Probability that EV $j$ selects CS $i$ at time $t$ \\
$x_{ij}(t)$ & Indicator: 1 if EV $j$ selects CS $i$ at $t$, else 0 \\
$\theta$ & Sensitivity parameter in MNL model \\
$\varepsilon_{ij}(t)$ & Perception error term (Gumbel-distributed) \\
$T_{ij}^{tra}(t)$ & Travel time from EV $j$ to CS $i$ \\
$T_R^{tra}(t)$ & Maximum reachable range based on SOC \\
$\widetilde{T_R^{tra}}(t)$ & Adjusted reachable range under risk and battery aging \\
$s_j^{ini}(t)$ & Initial SOC of EV $j$ at time $t$ \\
$Q_{j,\max}$ & Battery capacity of EV $j$ \\
$\varphi_j$ & Energy consumption rate of EV $j$ (km/kWh) \\
$\alpha_{\text{risk}}$ & Risk aversion coefficient of EV $j$ \\
$T$ & Vehicle age (years)\\
$\beta$ & Battery degradation rate \\
$T_{ij}^{c}(t)$ & Remaining charging time for EV $j$ at CS $i$ \\
$T_{ij}^{c,\text{ini}}(t)$ & Initial charging time before current visit \\
$Q_{ij}^{\Delta}(t)$ & Charging demand of EV $j$ at CS $i$ during $t$ \\
$f_i(\cdot)$ & Charging function of CS $i$ \\

\multicolumn{2}{l}{\textit{Station Operations and Queuing}} \\
$P_i(t)$ & Charging power (kW) of CS $i$ at time $t$ \\
$\lambda_i(t)$ & EV arrival rate at CS $i$ \\
$\mu_i$ & Service rate at CS $i$ \\
$S_i(t)$ & Number of charging poles at CS $i$ at time $t$ \\
$c_i$ & Total system capacity (charging + waiting spots) \\
$\pi_{i,d}(t)$ & Probability that CS $i$ has $d$ EVs in the system \\
$\pi_{i,0}(t)$ & Probability that CS $i$ is idle (no EVs) \\
$L_i(\lambda_i(t))$ & Expected queue length at CS $i$ \\
$W_i(\lambda_i(t))$ & Expected waiting time at CS $i$ \\
$T_{ij}^{q}(t)$ & Expected queuing time for EV $j$ at CS $i$ \\
$N_i^{\text{rejected}}(t)$ & Expected number of rejected EVs at CS $i$ \\

\multicolumn{2}{l}{\textit{Pricing and Economics}} \\
$Pr_i(t)$ & Charging price set by CS $i$ at time $t$ \\
$\nu_i(t)$ & Grid electricity price at CS $i$ \\
$\nu_{i,\text{max}}(t)$ & Maximum allowed price at CS $i$ \\
$R_{CS}$ & Total revenue from all CSs \\
$U_j$ & Utility of EV $j$ \\
$U_{EV}$ & Total utility of all EVs \\
$\eta_{\text{rejected}}$ & Penalty for rejected EVs \\
$PI$ & Total system performance index \\
$\omega$ & Weight for revenue vs. utility in objective \\
$\nu$ & Value of time for EV drivers \\
$\kappa$ & Marginal benefit of electricity consumption \\
$C_{ij}^{total}(t)$ & Total time-based cost from EV $j$ to CS $i$ \\

\multicolumn{2}{l}{\textit{Solution Algorithm (CEM and PSA)}} \\
$F(\boldsymbol{\theta})$ & Objective function for the optimization problem \\
$\boldsymbol{\theta}$ & Vector of decision variables (dynamic prices) \\
$\Omega$ & Feasible domain of the decision variables \\
$f(\boldsymbol{\theta}; \boldsymbol{\phi})$ & Probability density function of the sampling distribution \\
$\boldsymbol{\phi}^{(t)}$ & Parameters of the sampling distribution at iteration $t$ \\
$N$ & Population size of random samples in CEM \\
$q$ & Elite ratio for selecting best samples \\
$\hat{c}^{(t)}$ & Elite threshold value at iteration $t$ \\
$\beta$ (CEM) & Smoothing parameter for parameter update in CEM \\
$\epsilon$ & Convergence tolerance threshold \\
$D_{\text{KL}}$ & Kullback-Leibler divergence (Relative Entropy) \\
$\hat{f}^{(\ell)}(s)$ & Estimated PDF of the performance index at iteration $\ell$ \\
$\hat{g}_k^{(\ell)}(s)$ & Conditional PDF with variable $k$ fixed at its elite mean \\
$D_k^{(\ell)}$ & Probabilistic sensitivity index of variable $k$ at iteration $\ell$ \\
$\tau$ & Threshold for selecting active (sensitive) variables \\
$\mathcal{A}^{(\ell)}$ & Set of active (sensitive) decision variables at iteration $\ell$ \\
$\bar{\theta}_k^{(\ell)}$ & Mean of the elite samples for variable $k$ \\
$s_k^{(\ell)}$ & Standard deviation of the elite samples for variable $k$ \\
$\boldsymbol{\mu}^{(\ell)}$ & Mean vector of the sampling distribution at iteration $\ell$ \\
$\boldsymbol{\sigma}^{(\ell)}$ & Standard deviation vector of the sampling distribution at iteration $\ell$ \\

\end{longtable}

\subsection{The Formation of CS Attraction and EV Selection Model}

A set of fundamental assumptions is first established to lay the theoretical foundation. 
Subsequently, the CS attraction function, the MNL-based EV choice model, and their integration with queuing dynamics are developed step by step to describe EV-CS station selection behavior. 
Next, the reachability stations for the specific EVs, and on-road travel time, along with charging duration and energy demand, which are key components in evaluating the total charging cost, are introduced later. 
Overall, the proposed model explicitly accounts for both heterogeneity and stochasticity in EV users' decision-making processes.

\subsubsection{Assumptions}

\begin{assum}\label{assum:1}
Although various factors may influence EV charging decisions—such as socio-demographic characteristics and alternative-specific attributes—this study focuses on the operational attributes of CSs to develop a quantitatively tractable model of EV charging behavior.
\end{assum}

\begin{assum}\label{assum:2}
Each EV is assumed to have access to real-time information about the operational status of the charging system, including the number of occupied charging plugs and the expected number of incoming vehicles.
\end{assum}

\begin{assum}\label{assum:3}
Each EV makes its station selection only once, based on the real-time demand conditions at the moment of decision-making. Once a decision is made, the EV does not change its destination while en route to the selected station.
\end{assum}

\begin{assum}\label{assum:4}
EVs make decisions independently, considering only their own conditions and preferences. They do not observe or anticipate the real-time decisions of other EVs, introducing perceptual uncertainty and resulting in randomness in the utility evaluation of each charging station.
\end{assum}

\subsubsection{Modeling Charging Station Attractiveness}

To encourage EVs to utilize less congested CSs and thereby enhance overall system efficiency, it is essential to understand the underlying decision-making mechanism behind CS selection. Reilly's Law of Retail Gravitation, which explains how consumer demand is distributed among competing centers, provides a useful theoretical foundation for this modeling. By analogy, CSs that are closer and offer better service attributes—such as greater plug availability, lower prices, and shorter waiting times—are more likely to be chosen by EVs~\citep{Kazemtarghi2024}.

Building on Assumption~1, we propose an enhanced attraction function that incorporates additional influential factors, including charging power, dynamic pricing, and the total cost of the charging process:
\begin{equation}
Attr_{ij}(t)=\frac{S_i(t)\cdot P_i(t)}{Pr_i(t)\left(C_{ij}^{total}(t)\right)^2}
\end{equation}

Here, $Attr_{ij}(t)$ represents the time-varying attractiveness of CS $i$ to EV $j$; $S_i(t)$ denotes the number of available charging poles at CS $i$ during period $t$; and $P_i(t)$ is the available charging power (kW). $Pr_i(t)$ represents the dynamic charging price (\$/kWh), while $C_{ij}^{total}(t)$ denotes the total time-based cost incurred when EV $j$ travels to CS $i$, defined as:
\begin{equation}
C_{ij}^{total}(t)=T_{ij}^{tra}(t)+T_{ij}^{q}(t)+T_{ij}^{c}(t)
\end{equation}
where $T_{ij}^{tra}(t)$, $T_{ij}^{q}(t)$, and $T_{ij}^{c}(t)$ represent the travel, queuing, and charging times, respectively, all measured in hours (h).

\subsubsection{Modeling EV Choice via MNL}

To better capture the influence of incomplete information and bounded rationality on EV drivers' decisions, we introduce a random perception error term into the CS attractiveness formulation. The perception error is assumed to follow a Gumbel distribution. Based on the perceived attractiveness of CS $i$ for EV $j$, the selection probability is modeled via the multinomial logit (MNL) model as:
\begin{equation}
p_{ij}(t)=\frac{\exp\left(\theta\cdot Attr_{ij}(t)+\varepsilon_{ij}(t)\right)}{\sum\limits_{k\in \mathrm{ECA}_j(t)}\exp\left(\theta\cdot Attr_{kj}(t)+\varepsilon_{kj}(t)\right)}
\end{equation}

Here, $\theta\cdot Attr_{ij}(t)$ denotes the deterministic utility, and $\varepsilon_{ij}(t)$ represents the random perception error. Only CSs satisfying $T_{ij}^{tra}(t)\leq T_R^{tra}(t)$ are considered reachable, where $T_R^{tra}(t)$ represents the maximum reachable travel time. The set of reachable stations forms the Electricity Catchment Area (ECA) for EV $j$:
\begin{equation}
\mathrm{ECA}_j(t)=\{k\mid T_{kj}^{tra}(t)\leq T_R^{tra}(t)\}
\end{equation}

The probability that EV $j$ selects CS $i$ at time $t$ can thus be represented as a Bernoulli trial:
\begin{equation}
\mathbb{P}\left(x_{ij}(t)=1\right)=p_{ij}(t)
\end{equation}

By summing over all EVs, the expected number of arrivals at CS $i$ at time $t$ is:
\begin{equation}
\mathbb{E}[\lambda_i(t)]=\sum_j p_{ij}(t)
\end{equation}
where $\lambda_i(t)$ is the arrival rate of EVs at CS $i$. Hence, the aggregate arrival process follows from the sum of individual Bernoulli trials.

\subsubsection{Modeling Queuing Dynamics}

To capture the congestion effects on EV charging experiences, the service process at each CS is modeled using an M/M/$s$/$c$ queuing system. In this notation, the first ``M'' indicates that EV arrivals follow a Poisson process with rate $\lambda_i(t)$, while the second ``M'' denotes exponentially distributed service times with mean rate $\mu_i$. Furthermore, $s$ represents the number of active charging poles $S_i(t)$, and $c$ denotes the finite system capacity $c_i$, including both charging and waiting spaces~\citep{Shafiei2023}. 

If all charging poles are occupied when an EV arrives at a CS, the EV must wait in a queue. 
The charging process at a CS is represented as a Markov chain, where the system states are defined by the number of EVs in the system (including both charging and waiting vehicles), denoted by $d$. 
The Markov property holds since the number of EVs at any given time depends only on the previous system state. 
When the queue reaches its maximum length, it is constrained by the station capacity $c_i$, resulting in $c_i + 1$ possible states in total. 

Within each discrete time interval $t$, the arrival process follows a Poisson distribution with mean rate $\lambda_i(t)$, representing the expected number of EVs arriving to charge at the station. 
The service rate $\mu_i$ is conceptually associated with the active charging poles rather than being a strict inverse of charging duration. 
Although the number of active poles $S_i(t)$ may vary dynamically over time due to vehicles waiting from previous periods, the parameters $\lambda_i(t)$, $\mu_i$, and $S_i(t)$ are assumed constant within each time step. 
This assumption is made because, from an EV driver's perspective, the real-time charging demand of other users cannot be precisely predicted. 
Instead, each EV estimates its expected waiting time based on the known charging power and the current occupancy of the CS. 
Accordingly, the service rate is approximated as a constant value correlated with the charging power rather than its instantaneous real-time measure. 
The stationary probability distribution of the Markov chain, representing the probability that CS $i$ is in state $d$ at time $t$, can then be computed recursively as follows:

\begin{equation}
\pi_{i,d}(t)=
\begin{cases}
\dfrac{1}{d!}\left(\dfrac{\lambda_i(t)}{\mu_i}\right)^d\pi_{i,0}(t),& d\leq S_i(t)\\[6pt]
\dfrac{1}{S_i(t)!}\left(\dfrac{1}{S_i(t)}\right)^{d-S_i(t)}\left(\dfrac{\lambda_i(t)}{\mu_i}\right)^d\pi_{i,0}(t),& S_i(t)\leq d\leq c_i
\end{cases}
\end{equation}

where $\pi_{i,0}$ is the initial condition, and the probabilities of all system states satisfy the following constraint:
\begin{equation}
\sum_{d=0}^{c_i}\pi_{i,d}(t)=1
\end{equation}
Accordingly, $\pi_{i,0}(t)$ represents the probability that there is no vehicle in the queue at charging station $i$, which can be calculated as:
\begin{equation}
\pi_{i,0}(t)=
\left[
\sum_{d=0}^{S_i(t)-1}\frac{1}{d!}\left(\frac{\lambda_i(t)}{\mu_i}\right)^d
+\sum_{d=S_i(t)}^{c_i}\frac{1}{S_i(t)!}\left(\frac{1}{S_i(t)}\right)^{d-S_i(t)}\left(\frac{\lambda_i(t)}{\mu_i}\right)^d
\right]^{-1}
\end{equation}

Therefore, the relationship between arrival rate and expected queue length can be established directly:
\begin{equation}
L_i(\lambda_i(t))=\sum_{d=S_i(t)+1}^{c_i}(d-S_i(t))\pi_{i,d}(t)
\end{equation}

Then, according to Little's law, the average EV's waiting time at a CS $i$ can be calculated as follows:
\begin{equation}
T_{ij}^q(t)=W_i(\lambda_i(t))=\frac{L_i(\lambda_i(t))}{\lambda_i(t)(1-\pi_{i,c_i}(t))}
\end{equation}

Furthermore, if the expected number of EVs arriving at a particular CS far exceeds the station's capacity, then although these vehicles will be rejected for inclusion in the formal queue for average waiting time calculation, they still need to be considered as a source of charging system loss caused by congestion. Thus, the rejected EVs should be represented as:

\begin{equation}
N_i^{rejected}(t)=\lambda_i(t)\cdot \pi_{i,c_i}(t)
\end{equation}

\subsubsection{Modeling Reachability and Travel Time}

Before calculating the average waiting time for each EV, it is necessary to determine its departure time from the charging station. 
At this stage, the concept of the initial stage of charging (SOC) at the start of the charging event should be clarified. 
The departure time of an EV is inherently related to its initial SOC, which determines the maximum reachable range of the vehicle. 
For each EV $j$, the selection range must be smaller than the maximum range corresponding to its current SOC during period $t$, ensuring that the first-stage selection of charging station $C_i$ remains feasible. 

Accordingly, $T_R^{tra}(t)$ is defined as the maximum reachable range for EV $j$ at time $t$:
\begin{equation}
T_R^{tra}(t)=\frac{s_j^{ini}(t)\cdot Q_{j,\max}}{\varphi_j},
\end{equation}
where $s_j^{ini}(t)$ denotes the initial SOC of EV $j$, $Q_{j,\max}$ represents its maximum battery capacity, and $\varphi_j$ is the energy consumption rate (kWh/km), which may vary depending on the vehicle type. 
By borrowing the concept of catchment area commonly used in geography~\citep{Zhang2024,Chang2024}, the Electricity Catchment Area (ECA) is introduced to represent the set of charging stations that are reachable for a specific EV $j$ based on its current SOC. 
Accordingly, the ECA can be defined as:
\begin{equation}
\mathrm{ECA}(s_j^{ini}(t))=\{i\mid T_{ij}^{tra}(t)\leq T_R^{tra}(t)\}
\end{equation}
where $T_{ij}^{tra}(t)$ is the required travel distance from the EV’s current location to charging station $i$, and $T_{R}^{tra}(t)$ is the maximum reachable range defined in Equation 13.

\begin{equation}
\mathrm{ECA}(s_j^{ini}(t))=\{i\mid T_{ij}^{tra}(t)\leq T_R^{tra}(t)\}
\end{equation}


To reflect realistic driving conditions, we incorporate the Risk Aversion and Battery Deterioration (RABD) model:
\begin{equation}
\widetilde{T_R^{tra}}(t)=T_R^{tra}(t)\times(1-\alpha_{risk})\times e^{-\beta T}
\end{equation}
where $\alpha_{risk}$ represents the driver's risk aversion and $\beta$ denotes the rate of battery degradation. The adjusted ECA is defined as:
\begin{equation}
\widetilde{ECA}(s_j^{ini}(t))=\{i\mid T_{ij}^{tra}(t)\leq \widetilde{T_R^{tra}}(t)\}
\end{equation}

\subsubsection{Modeling Charging Duration and Demand}

In the preceding subsection, the initial SOC $s_j^{ini}(t)$, which reflects the partial charging prior to arrival and is related to the initial charging duration as follows:
\begin{equation}
s_j^{ini}(t) = \frac{f_i(T_{ij}^{c,ini}(t))}{Q_{j,\max}}.
\end{equation}

Owing to the nonlinear relationship between charging time and the obtained SOC, the time required to complete the final stage of charging (high SOC region) is typically much longer than that of the initial stage (low SOC region)~\citep{Wang2016}. Considering that the dynamic pricing model proposed in this study is developed under the time-based billing mechanism, it is essential to accurately characterize this nonlinear SOC-time relationship for Level 3 CSs to further describe the heterogeneity in modeling. 

Accordingly, the charging function $f_i(\cdot)$ follows the hierarchical model proposed by~\citep{Yu2021}, calibrated using the Tesla Model 3 (Long Range version) with a battery capacity of 75~kWh:
\begin{equation}
f_i(T)=
\begin{cases}
f_{L2}(T) = \varphi \cdot T, & C_i \in \mathbb{C}^{L2}, \\[6pt]
f_{L3}(T) = 1 + a e^{-bT} - (1 + a)e^{-cT}, & C_i \in \mathbb{C}^{L3},
\end{cases}
\end{equation}
where $T$ represents the charging time in minutes, and $f_i(T) \in [0,1]$ denotes the normalized SOC. 

For Level 2 (L2) charging, the SOC increases approximately linearly, corresponding to a constant charging power of about 20~kW. Given a battery capacity of 75~kWh, the constant charging rate is determined as $\varphi = 1/225$, implying that a full charge is achieved in approximately 225 minutes. 

For Level 3 (L3) DC fast charging, a bi-exponential model is adopted with parameters $a = 2.096$, $b = 0.0749$, and $c = 0.0552$. This nonlinear model accurately captures the typical charging characteristics, where the charging rate is initially high and gradually decreases as the battery approaches full capacity to mitigate degradation. The fitted L3 model satisfies the boundary conditions $f_{L3}(0) = 0$ and $\lim_{T \to \infty} f_{L3}(T) = 1$.


Therefore, the actual charging duration of vehicle $j$ at station $i$ can be expressed as:
\begin{equation}
T_{ij}^c(t) = f_i^{-1}(Q_j^{\max}) - T_{ij}^{c,ini}(t),
\end{equation}
where $f_i^{-1}(\cdot)$ denotes the inverse function of the charging curve $f_i(\cdot)$. 

Accordingly, the total charging demand over the entire charging process can be formulated as:
\begin{equation}
Q_{ij}^\Delta(t) = f_i\big(T_{ij}^{c,ini}(t) + T_{ij}^c(t)\big) - s_j^{ini}(t)\, Q_{j,\max}.
\end{equation}

\subsection{Game Formation}
\label{subsec:Game_Formation}

The proposed framework formulates a dynamic interaction and optimization process between EVs and CSs within a public charging network. The primary objective is to mitigate congestion-induced queuing losses and to improve resource allocation efficiency through dynamic pricing mechanisms. The resulting system is designed as a Stackelberg game, in which the government authority acts as the leader by setting time-varying prices, while EV users act as followers and respond by selecting their preferred Cs. Through this leader-follower interaction, the framework aims to achieve a system-level performance improvement that balances operator revenue and user welfare.

\subsubsection{Lower Level: EV Charging Choice and Queuing Loss Minimization}

At the lower level of the Stackelberg game, EV users are modeled as decentralized followers who independently select CSs based on their individual charging demands, perceived travel times, anticipated waiting times, and time-varying charging prices. Due to bounded rationality and incomplete information, EV charging choices are represented using a probabilistic MNL model, which captures stochastic perception errors in user decision-making.

\paragraph{Rejected EVs and Queuing Loss}

Each CS has a limited service capacity, and congestion may lead to the rejection of some EVs during peak demand periods. To characterize congestion-induced service failures, the expected number of rejected EVs at charging station \( i \) is approximated as
\begin{equation}
N_i^{\text{rejected}}(t) = \lambda_i(t)\,\pi_i^{\text{rej}}(t),
\end{equation}
where \( \lambda_i(t) \) denotes the EV arrival rate at station \( i \), and \( \pi_i^{\text{rej}}(t) \) represents the probability that an arriving EV is rejected due to insufficient available charging capacity, which is approximated based on station-level queuing conditions.

\paragraph{Individual EV Utility}

For each EV \( j \) that is successfully served by charging station \( i \), the perceived utility is defined as
\begin{equation}
U_j = \sum_t \sum_i \left[
\kappa Q_{ij}^{\Delta}(t)
- T_{ij}^{c}(t)\,Pr_i(t)
- T_{ij}^{q}(t)\,\nu
\right] p_{ij}(t),
\end{equation}
where \( Q_{ij}^{\Delta}(t) \) denotes the charged energy amount, \( T_{ij}^{c}(t) \) is the charging duration, \( Pr_i(t) \) is the charging price at station \( i \), \( T_{ij}^{q}(t) \) represents the expected waiting time, \( \nu \) is the unit waiting-time penalty, and \( p_{ij}(t) \) is the MNL-based probability that EV \( j \) selects charging station \( i \) at time \( t \).

\paragraph{Aggregate EV Utility}

The aggregate EV utility at the system level is defined as
\begin{equation}
U_{\text{EV}} =
\sum_j U_j
- \sum_t \sum_i N_i^{\text{rejected}}(t)\,\eta_{\text{rejected}},
\end{equation}
where the second term penalizes rejected EVs using a fixed penalty coefficient \( \eta_{\text{rejected}} \), reflecting the disutility associated with unmet charging demand due to congestion.

\subsubsection{Upper Level: Dynamic Pricing for System Performance Optimization}

At the upper level, the government authority acts as the leader and dynamically adjusts charging prices at each station over time to influence EV charging behavior at the lower level. Through these pricing decisions, the authority indirectly shapes station-level demand distributions and congestion patterns across the charging network.

To balance economic incentives and service quality, the upper-level objective incorporates both charging station revenue and EV user utility into a unified system performance index.

\paragraph{Charging Station Revenue}

Charging station revenue is primarily generated from electricity sales and is influenced by both dynamic prices and EV charging demand. The total revenue can be expressed as
\begin{equation}
R_{\text{CS}} =
\sum_t \sum_i \sum_j
Q_{ij}^{\Delta}(t)\,
\bigl(Pr_i(t) - \nu_i(t)\bigr)\,
p_{ij}(t),
\end{equation}
where \( \nu_i(t) \) denotes the wholesale electricity procurement price at charging station \( i \).

\paragraph{System Performance Index}

The overall system objective is formulated as a weighted sum of operator revenue and aggregate EV utility:
\begin{equation}
PI = \omega R_{\text{CS}} + (1 - \omega) U_{\text{EV}},
\end{equation}
where \( \omega \in [0,1] \) represents the relative importance assigned to charging station revenue versus EV user welfare. 
Based on the sensitivity analysis in~\ref{appendix:sensitivity_analysis}, a balanced setting (\( \omega = 0.5 \)) is adopted in this study.

\subsubsection{Constraints}

\begin{constraint}
The charging price is bounded by an upper and lower limit, where the lower limit corresponds to the grid electricity price, and the upper limit represents a predefined maximum allowable price:
\begin{align}
\nu_i(t) \le Pr_i(t) \le \nu_{i,\text{max}}(t)
\end{align}
\end{constraint}

\begin{constraint}
This constraint is defined in accordance with Equation~(17), which reflects that the remaining driving range of specific EVs is strictly bounded by the adjusted ECA considering risk aversion and battery deterioration.
\end{constraint}

\subsection{Game Solution}
As established in \cite{Jeroslow1985}, general bi-level programming problems are NP-hard, underscoring the inherent complexity of our model. To address the NP-hard stochastic optimization problem with queuing dynamics in our proposed Stackelberg game formulation, the Cross-Entropy Method (CEM), which is particularly suited for large-scale combinatorial optimization ~\citep{Tran2021b,Tran2023}, is proposed for upper-level dynamic pricing optimization, combined with the Method of Successive Averages (MSA) for lower-level EV-CS equilibrium analysis. In particular, we combine CEM with a dynamic Parameter Sensitivity Analysis (PSA) mechanism to mitigate the computational burden caused by the high-dimensional pricing strategy space. This hybrid approach accelerates the convergence of the upper-level optimization by adaptively identifying and focusing on the most influential station-hour price variables.

\subsubsection{The cross-entropy method}

As mentioned, a CEM-based approach has shown its superiority in finding a relatively good optimal solution in a range of large-scale combinatorial or continuous optimization problems~\citep{Tran2021b,Tran2023,ngoduy2013,Zhong2016}. The CEM algorithm is thus proposed to solve this dynamic pricing problem and can be broken down into two key steps:
\begin{enumerate}
    \item Generate a number of trial parameter sets randomly according to the chosen distributions.
    \item Based on the values of the objective function associated with each trial parameter set, update the probability distribution used to generate the random trial sets according to the principle of ``importance sampling''.
\end{enumerate}
For a general optimization problem, obtaining a global optimum can be regarded as a rare event. 
The Cross-Entropy Method (CEM) provides a Monte Carlo-based framework to estimate such rare-event probabilities and reformulates the optimization process into a probabilistic learning problem. The detailed derivation of the CEM used in this study is provided in~\ref{appendix:cem}.

\subsubsection{Probabilistic sensitivity analysis}

However, the optimization of stochastic systems such as queue dynamics or EV charging networks often involves parameters with uncertain or correlated effects on system performance and also invloving the high-dimensional decision space may lead to slow convergence of the CEM algorithm. To address these challenges, \citet{Zhong2016} introduced a PSA-CEM framework based on information theory, which quantifies how variations in input parameters affect the probability distribution of model outputs \citep{Liu2006}, to provide a more general measure of output variability, and further incorperating the PSA into the CEM sampling procedures to ``masked'' smaller parameters to reduce search space for the high-dimensional decision space.

Instead of using conventional variance-based indices ~\citep{Saltelli2008}, PSA started to employ the relative entropy (i.e., Kullback-Leibler divergence) between the unconditional and conditional probability density functions (PDFs) of the system performance measure. This entropy-based index provides a robust measure of sensitivity even when the distributions are nonlinear, multimodal, or heavy-tailed. Specifically, given two distributions, the unconditional $f(y)$ and the conditional $g(y|x_i)$ when parameter $x_i$ is fixed, the relative entropy is expressed as:
\begin{equation}
D_{\text{KL}}(g(y|x_i)\,\|\,f(y)) = 
\int g(y|x_i)\ln\!\frac{g(y|x_i)}{f(y)}\,dy,
\label{eq:kl_divergence}
\end{equation}
where a larger $D_{\text{KL}}$ indicates that fixing $x_i$ produces a greater deviation in the output PDF, 
and hence $x_i$ is more influential.

This ﬂexibility facilitates the usage of the relative entropy based PSA under various scenarios of design under uncertainty, e.g. reliability-based design, robust design, and utility optimization. It is claimed that the relative entropy based PSA can be applied to both the prior-design stage for variable screening when a design solution is yet identiﬁed and the post-design stage for uncertainty reduction after an optimal design has been determined \citep{Liu2006}.

\subsubsection{Dynamic probabilistic sensitivity analysis and adaptive CEM search}
\label{subsec:dynamic_PSA_CEM}

Building upon the relative entropy-based PSA proposed by \citet{Liu2006} and its application to car-following model calibration in \citet{Zhong2016}, this study develops a dynamic PSA-guided CEM framework. In contrast to the static preconditioning in previous work, where PSA is performed once prior to optimization and the set of important variables is fixed throughout, the proposed scheme updates the PSA indices periodically during the CEM iterations based on the most recent sample population. 

This adaptive mechanism allows the algorithm to re-evaluate, in an online fashion, the influence of each station-hour price component on the performance index and to dynamically distinguish sensitive decision variables from insensitive ones.

Let $\boldsymbol{\theta} \in \mathbb{R}^{d}$ denote the stacked decision vector of all station-hour prices
\[
\boldsymbol{\theta}
=
\big(
p_{11},\ldots,p_{1N_h},
\ldots,
p_{N_s1},\ldots,p_{N_sN_h}
\big)^\top,
\]
where $N_s$ is the number of charging stations and $N_h$ is the number of hourly periods, so that $d = N_s N_h$.  
The performance index (objective function) associated with a given price vector $\boldsymbol{\theta}$ is denoted by $F(\boldsymbol{\theta})$, which aggregates the EV and CS utilities as defined in Section~\ref{subsec:Game_Formation}.

At CEM iteration $\ell$, a population of $N$ price samples $\{\boldsymbol{\theta}^{(m)}\}_{m=1}^N$ is drawn from the current sampling distribution $f^{(\ell)}(\boldsymbol{\theta})$, and the corresponding scores $\{F^{(m)}\}_{m=1}^N$ are evaluated, where $F^{(m)} = F(\boldsymbol{\theta}^{(m)})$. 
Using these samples, we estimate the PDF $\hat{f}^{(\ell)}(f)$ of the performance index $F$ approximated by a Gaussian distribution.  
For each decision variable $\theta_k$ (i.e., a specific station-hour price), we construct an associated "frozen" population by fixing $\theta_k$ at its elite mean $\bar{\theta}_k^{(\ell)}$ while keeping the remaining components unchanged, and obtain a corresponding normal density estimate $\hat{g}_k^{(\ell)}(f)$ of the performance index. 
Following the above-mentioned \eqref{eq:kl_divergence}, the relative-entropy-based probabilistic sensitivity index of $\theta_k$ at iteration $\ell$ is defined as:
\begin{equation}
D_k^{(\ell)}
=
\int
\hat{g}_k^{(\ell)}(s)
\ln
\frac{\hat{g}_k^{(\ell)}(s)}{\hat{f}^{(\ell)}(s)}
\, \mathrm{d}s,
\label{eq:dynamic_PSA_index}
\end{equation}
which quantifies the change in the distribution of the performance index when the variability in $\theta_k$ is removed. 
And then, deﬁne a simpliﬁed problem that considers only the most sensitive decision variables by introducing a user-specified threshold $\tau>0$, an active set of sensitive decision variables at iteration $\ell$ is then defined as
\begin{equation}
\mathcal{A}^{(\ell)}
=
\left\{
k \in \{1,\ldots,d\} \; \big| \; D_k^{(\ell)} > \tau
\right\}.
\label{eq:active_set}
\end{equation}
Only the variables in $\mathcal{A}^{(\ell)}$ are allowed to evolve according to the standard CEM update rule, whereas the remaining variables are temporarily frozen around their current means and variances. 
Let $\boldsymbol{\mu}^{(\ell)}$ and $\boldsymbol{\sigma}^{(\ell)}$ denote the mean vector and standard deviation vector of the sampling distribution at iteration $\ell$, and let $\bar{\boldsymbol{\theta}}^{(\ell)}$ and $\boldsymbol{s}^{(\ell)}$ denote the sample mean and sample standard deviation of the elite set selected at iteration $\ell$. 
The adaptive PSA-guided CEM update is given component-wise by:
\begin{equation}
\mu_k^{(\ell+1)} =
\begin{cases}
\beta \mu_k^{(\ell)} + (1-\beta)\,\bar{\theta}_k^{(\ell)}, 
& \text{if } k \in \mathcal{A}^{(\ell)},\\[0.2em]
\mu_k^{(\ell)}, 
& \text{if } k \notin \mathcal{A}^{(\ell)},
\end{cases}
\label{eq:mu_update_dynamic}
\end{equation}
\begin{equation}
\sigma_k^{(\ell+1)} =
\begin{cases}
\beta \sigma_k^{(\ell)} + (1-\beta)\,s_k^{(\ell)}, 
& \text{if } k \in \mathcal{A}^{(\ell)},\\[0.2em]
\sigma_k^{(\ell)}, 
& \text{if } k \notin \mathcal{A}^{(\ell)},
\end{cases}
\label{eq:sigma_update_dynamic}
\end{equation}
where $\beta \in [0,1]$ is the smoothing parameter that controls the degree of update between iterations. According to empirical findings in the CEM literature, $\beta$ values between 0.4 and 0.9 generally yield satisfactory results; in this study, $\beta$ is set to 0.7.

Overall, this adaptive ``freezing-unfreezing'' strategy effectively realizes an on-line variable screening within the CEM framework. Decision variables with persistently low probabilistic sensitivity are kept close to their baseline values, whereas highly sensitive variables continue to evolve toward the Pareto-optimal region. 
As a result, the proposed dynamic PSA-guided CEM is able to balance exploration and exploitation more efficiently, reduce the computational burden in high-dimensional price spaces, and alleviate premature convergence, thus extending the static PSA-CEM formulation of \citet{Zhong2016} to a fully adaptive setting.


\subsubsection{The Hybrid CEM-MSA optimization framework}
Building on the dynamic PSA-CEM model for solving the upper-level dynamic pricing problem, the overall framework also employs the Method of Successive Averages (MSA) at the lower level to compute the EV-CS choice equilibrium and capture queue dynamics.

To address the computational challenges and the large solution space, our framework incorporates three key features:

(1) Rolling-horizon strategy.
The entire simulation horizon is divided into several hourly optimization windows to dynamically capture the temporal evolution of queuing phenomena and charging behaviors, while maintaining computational tractability.

(2) Dynamic PSA-CEM mechanism.
As detailed in Section \ref{subsec:dynamic_PSA_CEM} and Algorithm \ref{alg:cem-msa}, the proposed dynamic PSA-guided CEM adaptively identifies critical decision variables during optimization.
It is particularly effective for large-scale combinatorial problems and demonstrates strong robustness and superiority in approaching near-global optimal solutions \citep{rubinstein2004crossentropy}.

\begin{figure}[htbp]
    \centering
    \includegraphics[width=1\textwidth]{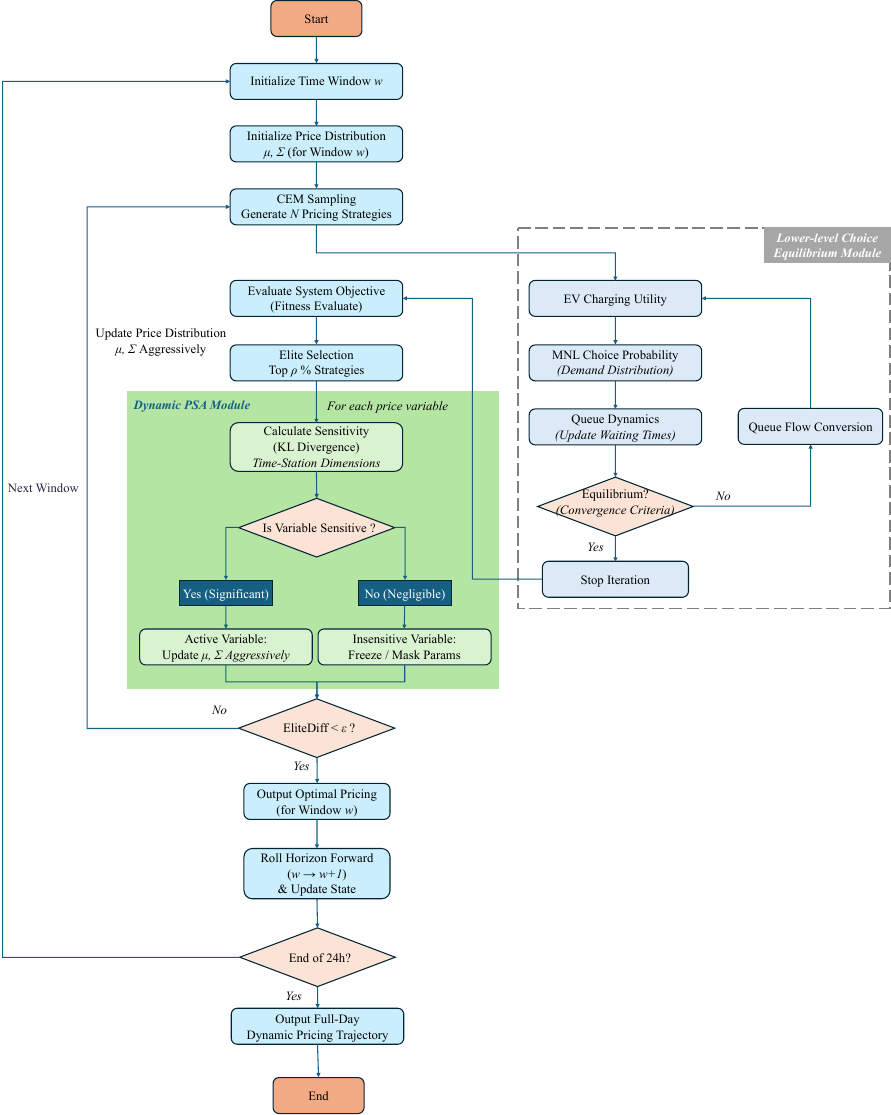}
    
    \caption{The overall algorithmic structure of the proposed bi-level optimization framework.}
    
    \label{fig:optimization_workflow}
\end{figure}

(3) Algorithmic structure design.
The overall framework draws inspiration from prior studies that combine hybrid metaheuristic algorithms with MSA for solving bi-level optimization problems under stochastic conditions \citep{Tran2021a,Tran2021b}.
At the lower level, this structure effectively captures the equilibrium charging decisions of EV users under incomplete information.
At the upper level, it adaptively optimizes the dynamic pricing strategy, improving both system-wide efficiency and operational stability.
Together, these features highlight the practicality and applicability of integrating user-equilibrium modeling with upper-level dynamic pricing optimization in real-world mobility systems, as the optimization workflow illustrated in Figure~\ref{fig:optimization_workflow} demonstrates.

\begin{algorithm}[t]
\caption{Dynamic PSA-guided CEM-MSA algorithm for rolling-horizon dynamic pricing optimization}
\label{alg:cem-msa}
\KwIn{Network topology, EV and CS parameters, $\kappa$, $\varphi$, $Q$, $\eta$, $\mu$, $T$, $H$, $\omega$, sample size $N$, PSA threshold $\tau$, smoothing factor $\beta$, PSA frequency $\psi$}
\KwOut{Optimized pricing profile $Pr^{*}$, user equilibrium flows $p^{*}$, performance index $PI^{*}$}

Initialize sampling distribution $\gamma^{(0)} = \mathcal{N}(\mu^{(0)}, \sigma^{(0)})$, iteration $\ell \gets 0$\;
\For{each rolling window $[t, t+H-1]$ over horizon $T$}{
  Initialize queue states and initial prices $Pr^{(0)}(t)$\;
  \While{stopping criterion not satisfied}{
    Draw $N$ pricing samples $\{Pr^{(m)}(t)\}$ from $f^{(\ell)}(\boldsymbol{\theta})$\;
    \For{each sample $Pr^{(m)}(t)$}{
      Compute $ECA^{(m)}(t)$, derive $p^{(0,m)}(t)=\text{MNL}(ECA^{(m)}(t))$, and update $p^{(m)}(t)=\text{MSA}(p^{(0,m)}(t))$\;
      Evaluate queue delay $T^q_{ij}(t)$, queue loss $QL^{(m)}(t)$, and utilities $(U_{\text{EV}}^{(m)}, R_{\text{CS}}^{(m)})$\;
      Compute performance $F^{(m)}(t)=\omega R_{\text{CS}}^{(m)}+(1-\omega)U_{\text{EV}}^{(m)}$\;
    }
    Select top $N_e$ elite samples and estimate $\hat{f}^{(\ell)}(s)$\;

    \If{mod($\ell$, $\psi$)$=0$}{
      \tcp{Conditionally triggered Dynamic PSA update}
      \For{each decision variable $\theta_k$}{
        Construct frozen population and estimate $\hat{g}_k^{(\ell)}(s)$\;
        Compute sensitivity $D_k^{(\ell)} = \int \hat{g}_k^{(\ell)}(s)\ln[\hat{g}_k^{(\ell)}(s)/\hat{f}^{(\ell)}(s)]\,ds$\;
      }
      Determine active set $\mathcal{A}^{(\ell)}=\{k|D_k^{(\ell)}>\tau\}$\;
      \For{each variable $k$}{
        \eIf{$k \in \mathcal{A}^{(\ell)}$}{
          $\mu_k^{(\ell+1)}=\beta\mu_k^{(\ell)}+(1-\beta)\bar{\theta}_k^{(\ell)}$\;
          $\sigma_k^{(\ell+1)}=\beta\sigma_k^{(\ell)}+(1-\beta)s_k^{(\ell)}$\;
        }{
          Keep frozen: $\mu_k^{(\ell+1)}=\mu_k^{(\ell)}$, $\sigma_k^{(\ell+1)}=\sigma_k^{(\ell)}$\;
        }
      }
    }
    \Else{
      \tcp{Standard CEM update (no Dynamic PSA this iteration)}
      Update all $\mu_k^{(\ell+1)}$, $\sigma_k^{(\ell+1)}$ using elite means $\bar{\theta}_k^{(\ell)}$ and deviations $s_k^{(\ell)}$\;
    }
    $\ell \gets \ell + 1$\;
  }
  Store optimized $Pr^{*}(t)$ and equilibrium $p^{*}(t)$\;
}
\Return $Pr^{*}, p^{*}, PI^{*}$\;
\end{algorithm}

The detailed solution procedure, combining CEM and MSA within a rolling-horizon framework, is presented in Algorithm~\ref{alg:cem-msa}. Specifically, we adopt a sample size of $N = 1000$ and set the elite sample proportion to $\rho = 1\%$. The parameter distribution is updated at each iteration using a smoothing rate $\alpha = 0.7$. The convergence criterion is defined such that the relative difference between the maximum and minimum utility values among elite samples remains below a predefined threshold $\varepsilon = 10^{-3}$ for two consecutive iterations.

\section{Experiment results}
\subsection{Experimental Setup}  
This study investigates the spatial distribution and temporal demand of electric vehicle (EV) charging infrastructure in the Clayton area of Melbourne. CS data were collected from \texttt{www.plugshare.com}, with stations categorized into two types: Fast-Charging Stations (FCS) equipped with CCS2 connectors, and Level 2 stations featuring Type 2 or J-1772 connectors. The study area comprises a total of 22 stations, including 19 FCSs and 3 Level 2 stations.  

\begin{figure}[htbp]
\centering
\includegraphics[width=1.0\linewidth]{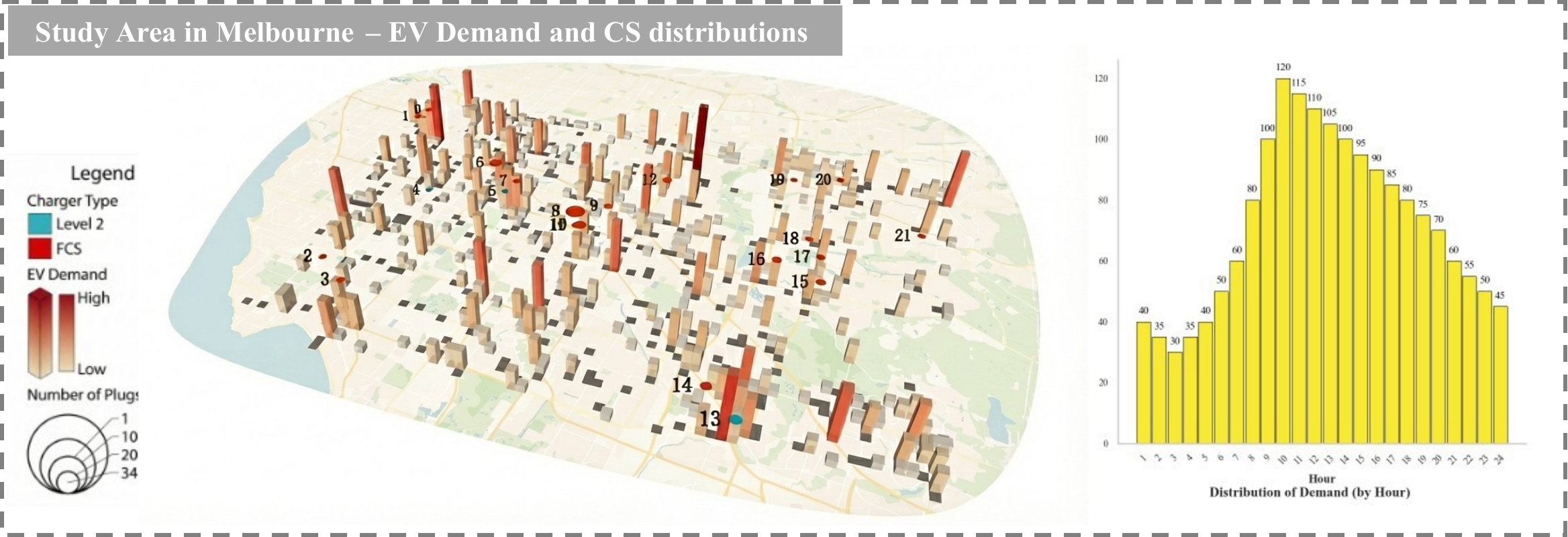}
\caption{Spatial and temporal overview of the study area: (Left) Spatial distribution of Level 2 and Fast-Charging Stations (FCS) with EV demand hotspots in Melbourne; (Right) Hourly EV charging demand profile.}
\label{fig:study_area_demand}
\end{figure}

The transport network is represented by major road segments within the study boundary, serving as potential EV trajectories from the well-known Openstreetmap dataset. Temporal charging demand profiles are derived from a real-world Australian dataset \citep{Islam2016}, exhibiting distinct peak (8:00 a.m.-5:00 p.m.) and off-peak periods. As shown in Figure~\ref{fig:study_area_demand}, the spatial distribution of CSs and the hourly demand patterns form the basis of the simulation environment.  

In the simulation, EVs are initially placed randomly within the road network, with optional clustering in demand hotspots for realism. Initial state-of-charge (SOC) values, battery capacities, vehicle ages, and drivers' risk attitudes are assigned according to pre-defined distributions. The energy consumption rate and charging process follow standard EV specifications and empirical charging behavior \citep{Yu2021,Islam2016}, incorporating power-time tapering effects. Consequently, the required charging duration for each EV depends on both its remaining battery capacity and the power rating of the selected CS.  

\begin{table}[H]
\centering
\caption{Key simulation and algorithm parameters used in the study.}
\label{tab:sim_params}
\renewcommand{\arraystretch}{1.2}
\begin{tabular}{lll}
\hline
\textbf{Parameter} & \textbf{Symbol} & \textbf{Value} \\
\hline
\multicolumn{3}{l}{\textit{EV and Economic Utility Parameters}} \\
Energy Consumption & $\varphi$ & 5.0 km/kWh \\
Charging Satisfaction Factor & $\kappa$ & 1.0 \$/kWh \\
Waiting Time Penalty (VOT) & $\nu$ & 5.0 \$/h \\
System Rejection Penalty & $\eta_{\text{rejected}}$ & 30.0 \$/EV \\
Risk Aversion Coefficients & $\alpha_{\text{risk}}$ & $\{0.0, 0.05, 0.15\}$ \\
\hline
\multicolumn{3}{l}{\textit{Pricing Constraints}} \\
Grid Base Price & $\nu_{i}(t)$ & 0.20 \$/kWh \\
Minimum Price Limit & $Pr_{\min}$ & 0.20 \$/kWh \\
Maximum Price Cap & $Pr_{\max}$ & 0.80 \$/kWh \\
\hline
\multicolumn{3}{l}{\textit{Algorithm (PSA-CEM-MSA)}} \\
Rolling Horizon Window & $H$ & 1 hour \\
Sample Size & $N$ & 1000 \\
Elite Ratio & $\rho$ & 0.05 (top 5\%) \\
Smoothing Factor & $\alpha$ & 0.7 \\
PSA Update Frequency & $\psi$ & 5 iterations \\
Convergence Threshold & $\epsilon$ & $10^{-3}$ \\
\hline
\end{tabular}
\end{table}

Key system parameters, EV attributes, and algorithm hyperparameters used in the simulations are summarized in Table~\ref{tab:sim_params}. All experiments were implemented in Python~3.9 and executed on a high-performance computer equipped with an Intel(R) Ultra~9~185H CPU (2.30~GHz) and 32~GB RAM.

\subsection{Convergence of the proposed algorithm}

\begin{figure}[htbp]
  \centering
  \includegraphics[width=0.9\linewidth]{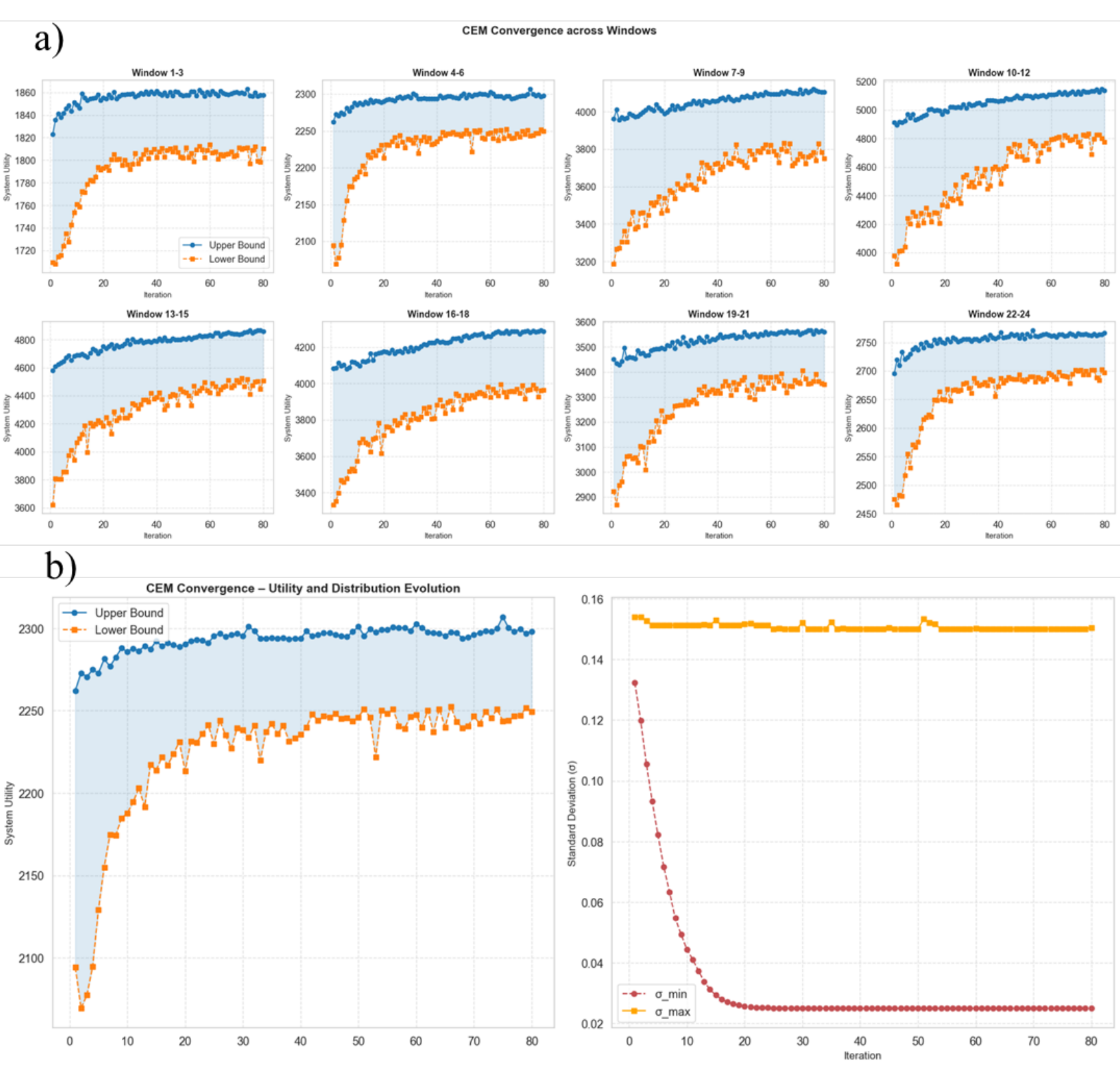}
  
  \caption{Convergence analysis of the CEM algorithm. (a) System utility bounds across 24 time windows, highlighting algorithmic robustness. (b) Detailed evolution of utility and standard deviation ($\sigma$), illustrating search space contraction.}
  \label{fig:cem_convergence}
\end{figure}

To validate the convergence of the proposed algorithm, its performance is examined over a 24-hour simulation horizon divided into hourly optimization windows. The convergence behaviors of the upper and lower bounds of the overall system utility across these time windows are illustrated in Figure~\ref{fig:cem_convergence}. Overall, the results exhibit a clear upward convergence trend, indicating that the proposed algorithm consistently improves system performance over successive iterations. Notably, the gap between the upper and lower bounds gradually narrows as iterations proceed, demonstrating the algorithm’s capability to progressively refine pricing strategies and EV-charging station (EV-CS) matching decisions.

Across all time windows, the lower bound of system utility increases rapidly within the first 30 iterations during periods of relatively low charging demand, particularly at the beginning and end of the day. In contrast, during peak demand periods (e.g., from 7:00 a.m. to 6:00 p.m.), the convergence process becomes more gradual. This behavior reflects the increased complexity of balancing supply and demand under higher congestion levels, with convergence typically occurring after more than 60 iterations.

To further demonstrate the optimality, robustness, and convergence characteristics of the proposed PSA-CEM algorithm, Figure~\ref{fig:cem_convergence}(b) presents the detailed evolution of system utility and the associated standard deviation bounds, namely $\sigma_{\min}$ and $\sigma_{\max}$, for the second time window over the 24-hour horizon. Specifically, $\sigma_{\max}$ represents the initial exploration variance that maintains sufficient diversity in candidate pricing solutions, while $\sigma_{\min}$ defines a lower bound that prevents premature convergence.

As illustrated, the system utility exhibits a steady upward trend, while the effective standard deviation progressively decreases from $\sigma_{\max}$ toward $\sigma_{\min}$. This behavior indicates that the algorithm gradually shifts from global exploration to local exploitation, thereby contracting the search space in a controlled manner. Once the standard deviation approaches $\sigma_{\min}$, the pricing distribution becomes highly concentrated, suggesting that the algorithm has converged to a stable and near-optimal pricing strategy for the charging station. The bounded variance mechanism governed by $\sigma_{\min}$ and $\sigma_{\max}$ ensures both convergence stability and solution robustness under varying demand levels, explaining the smooth yet reliable convergence patterns observed across different time windows.

\begin{figure}[htbp]
    \centering
    \includegraphics[width=1\linewidth]{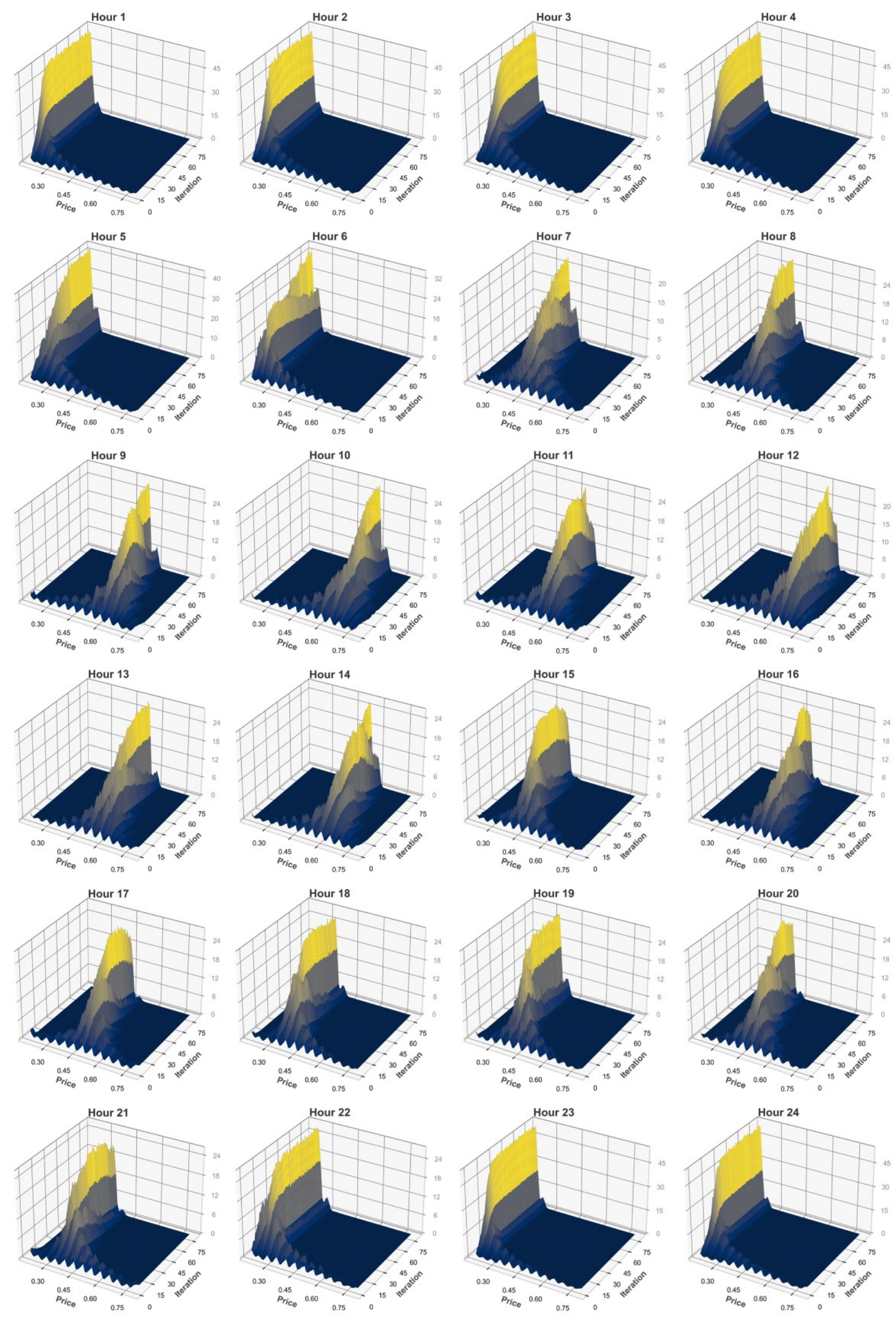}
    \caption{3D PDF evolution of price distributions for Station 8 across a 24-hour horizon, illustrating the convergence of price density over optimization iterations.}
    \label{fig:price_3d_evolution}
\end{figure}

Furthermore, to provide a more intuitive illustration of the convergence behavior of pricing strategies, Station~8 is selected as a representative case study. Figure~\ref{fig:price_3d_evolution} depicts the evolution of the three-dimensional probability density function (PDF) of charging prices at Station~8 over the 24-hour horizon. Consistent with typical CEM convergence characteristics, the price distribution is initially broad, reflecting high uncertainty in optimal pricing. As iterations progress, the distribution gradually narrows and converges toward a stable pricing range, indicating that the CEM-MSA algorithm effectively identifies optimal prices that balance charging demand and congestion effects.

Although all time windows exhibit eventual convergence, the convergence patterns differ across time periods due to variations in demand levels and charging station layouts. During peak hours, the price distribution first contracts toward a relatively narrow region, then slightly widens before finally stabilizing at the optimal price range. This phenomenon reflects the complex interplay among high demand, congestion, and pricing adjustments. In contrast, during off-peak hours, the price distribution converges more directly to a stable range, indicating lower variability in optimal pricing under reduced congestion conditions. These observations highlight the adaptive capability of the proposed algorithm in responding to temporal demand fluctuations.

Finally, from the perspective of price magnitude, the 3D PDF evolution reveals a clear yet non-monotonic pricing pattern at Station~8 over the 24-hour horizon. Specifically, charging prices start at a relatively low level during the early hours, then increase with rising demand before the 10th hour. Prices remain around 0.5 AUD/kWh until approximately the 15th hour to 0.4 AUD/kWh, followed by a short-term decrease, and stabilize again around 0.5 AUD/kWh during the 16th and 17th hours. Subsequently, prices gradually decline toward the lowest level of approximately 0.2 AUD/kWh by the end of the day.

Importantly, this complex and non-monotonic pricing trajectory—rather than a simple monotonically increasing or decreasing trend—further demonstrates the realism of the pricing outcomes and underscores the robustness of the proposed algorithm in capturing dynamic demand-congestion interactions.

\subsection{Comparative Analysis of EV-CS Pairing Strategies under Stochastic Choice}

This section evaluates the performance of the proposed optimization algorithm by comparing three distinct CS selection strategies: (1) CEM-DC (Direct Choice, Deterministic), (2) CEM-Standard (MNL-based Stochastic Choice without MSA), and (3) CEM-MSA (The proposed MNL-MSA Equilibrium Choice). The comparison highlights how integrating stochasticity with Mean Successive Averages (MSA) impacts both system-wide utility and individual EV decision-making.

\subsubsection{Macro-Level Analysis: Convergence of System Utility}

\begin{figure}[htbp]
    \centering
    \includegraphics[width=1.0\linewidth]{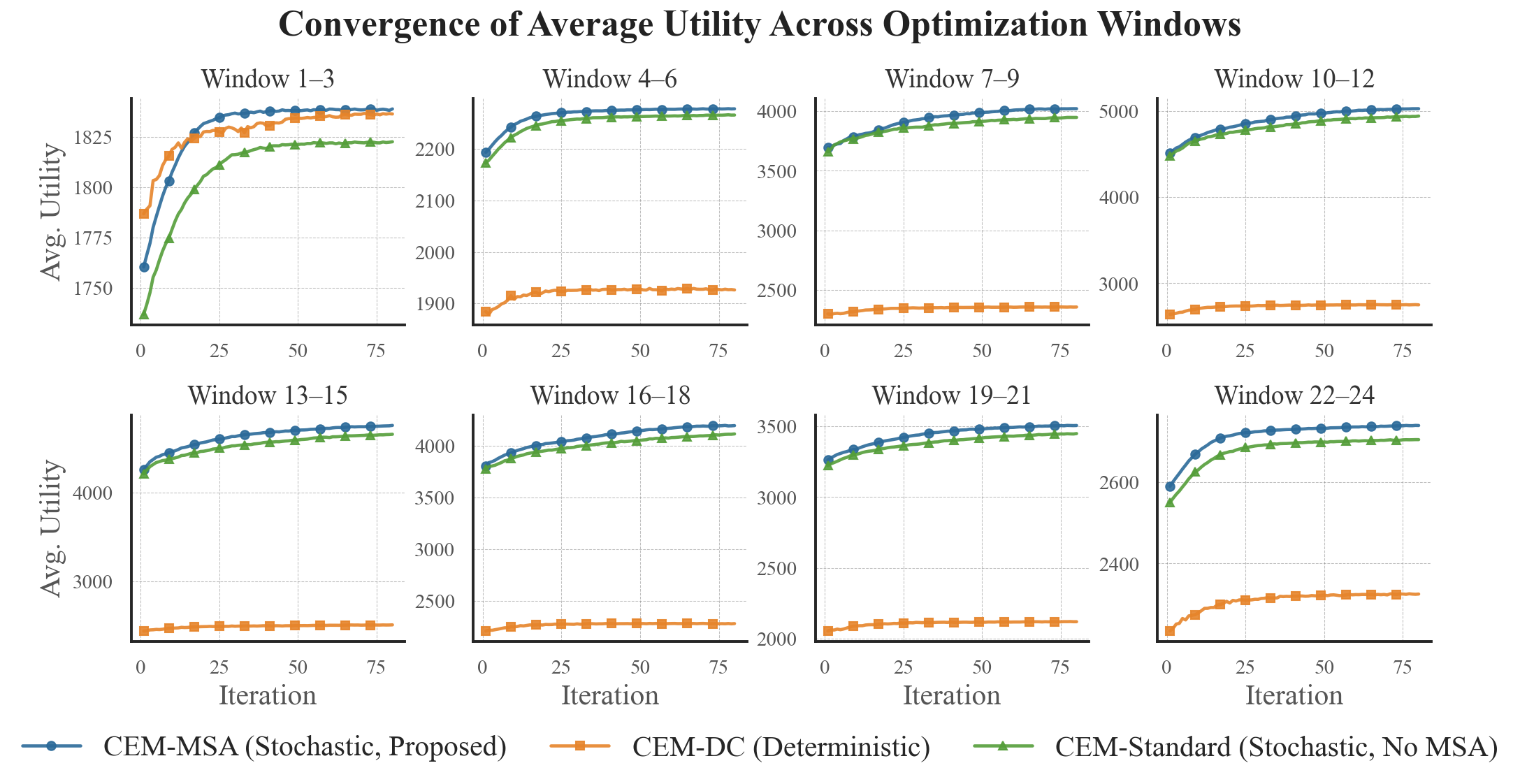}
    \caption{Convergence of average system utility across eight optimization window groups, comparing the proposed CEM-MSA against CEM-DC and CEM-Standard baselines.}
    \label{fig:algo_convergence_comparison}
\end{figure}

The first dimension of analysis focuses on the convergence of average system utility across different optimization windows, as illustrated in Figure~\ref{fig:algo_convergence_comparison}.

As observed in the figure, the convergence trends vary significantly across the strategies. The CEM-DC strategy (orange line) converges rapidly but stabilizes at a suboptimal utility level. While DC performs adequately in early windows (e.g., Window 1-3) where system congestion is lower, its inability to account for stochastic user behavior limits its potential for system-wide optimization. Conversely, the CEM-Standard strategy (green line), which introduces MNL-based randomness without equilibration, exhibits the lowest performance during the initial phases. This suggests that merely introducing stochasticity without a mechanism to balance demand initially introduces noise that degrades efficiency. However, as system congestion increases in subsequent windows, CEM-Standard overtakes the deterministic baseline, securing the second rank and maintaining it through to the end. This shift underscores the inherent value of stochasticity in mitigating the rigidity of deterministic choices as system complexity grows.

Notably, the proposed CEM-MSA (blue line) consistently achieves the highest average utility across all time horizons. By incorporating both the realistic stochasticity of EV choices and the MSA algorithm to reach equilibrium, CEM-MSA effectively mitigates congestion effects. It outperforms the deterministic baseline even in uncongested scenarios and demonstrates superior scalability as the optimization window progresses.

\subsubsection{Micro-Level Analysis: EV Choice Probability Distribution}

\begin{figure}[htbp]
    \centering
    \includegraphics[width=1.0\linewidth]{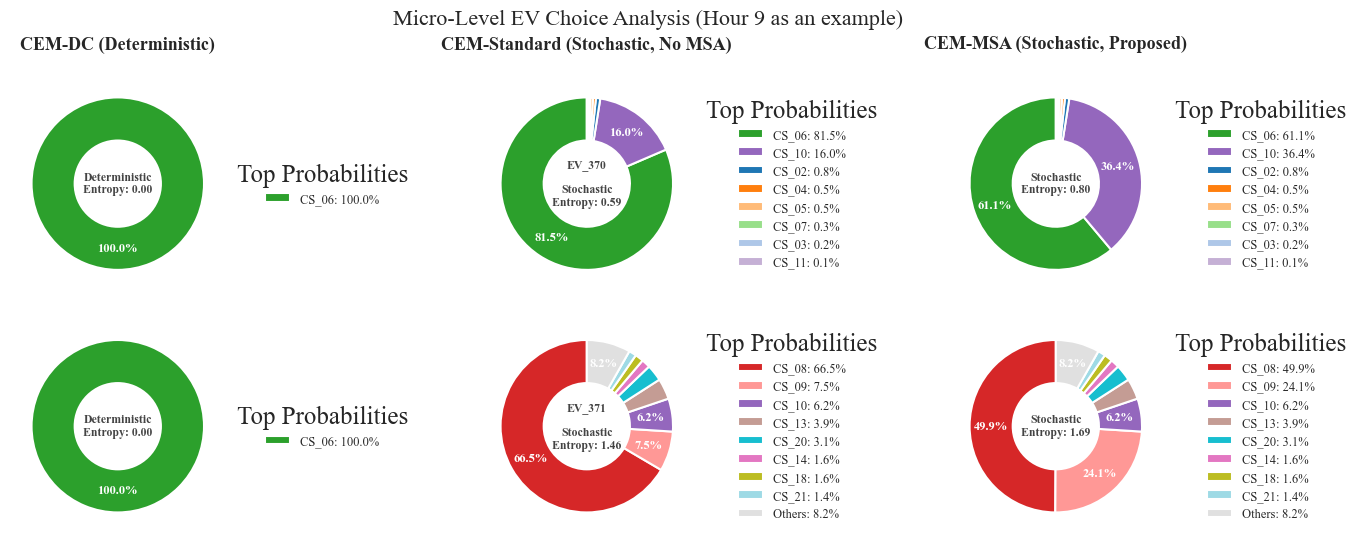}
    \caption{Micro-level EV choice analysis comparing CEM-DC (Deterministic), CEM-Standard (Stochastic), and the proposed CEM-MSA (Stochastic with MSA) for representative EVs at Hour 9.}
    \label{fig:ev_choice_analysis}
\end{figure}

To understand the mechanics driving these utility gains, we extend the analysis to the micro-level EV choice behaviors, as presented in Figure~\ref{fig:ev_choice_analysis}. This figure compares the choice probability distributions for two representative vehicles (EV 370 and EV 371) at Hour 9.

Under the CEM-DC criterion, the choice is binary and deterministic. Both EVs identify CS 06 as the optimal station and target it with 100\% probability (Entropy: 0.00). While this appears efficient individually, it leads to severe overcrowding at CS 06 when aggregated across the fleet.

In contrast, the stochastic methods distribute the demand. The CEM-Standard approach disperses the choice probability across a wide range of stations (e.g., EV 371 considers over 9 stations). While this leads to high entropy (e.g., 0.99 for EV 370) and distributes demand more broadly than the deterministic approach, the allocation remains potentially inefficient. This is because the dispersion is based solely on perception error and lacks system-level awareness (coordination), resulting in random rather than strategic distribution.

The CEM-MSA strategy strikes a critical balance. As shown in the third column of Figure~\ref{fig:ev_choice_analysis}, MSA modulates the probabilities found in the Standard model. For instance, for EV 370, CEM-MSA reduces the dominance of the primary choice (CS 06) from 81.5\% (Standard) to 61.1\%, while shifting significant probability (36.4\%) to a secondary option (CS 10). This shift demonstrates the equilibrium effect: the algorithm identifies that CS 06 is becoming over-congested and guides a portion of demand toward suboptimal but under-utilized stations. This redistribution guarantees that EVs avoid the "best" option when it is congested, leading to a higher global system optimum.

\subsection{Heterogeneity Analysis of Charging Stations}

Heterogeneity plays a critical role in shaping charging demand distribution, pricing outcomes, and user-perceived utility in EV charging systems. In this subsection, heterogeneity is examined under the converged solution of the proposed CEM-MSA framework. Specifically, we conduct an ex-post analysis to investigate how different station-level characteristics influence equilibrium traffic loads, dynamic pricing strategies, and EV user utility, without introducing additional decision variables into the optimization process. Two major sources of heterogeneity are considered: charger type heterogeneity, which affects service rates, and capacity heterogeneity, which influences the scale of charging demand aggregation.

\subsubsection{Charger Type Heterogeneity}

\begin{figure}[htbp]
    \centering
    \includegraphics[width=1.0\linewidth]{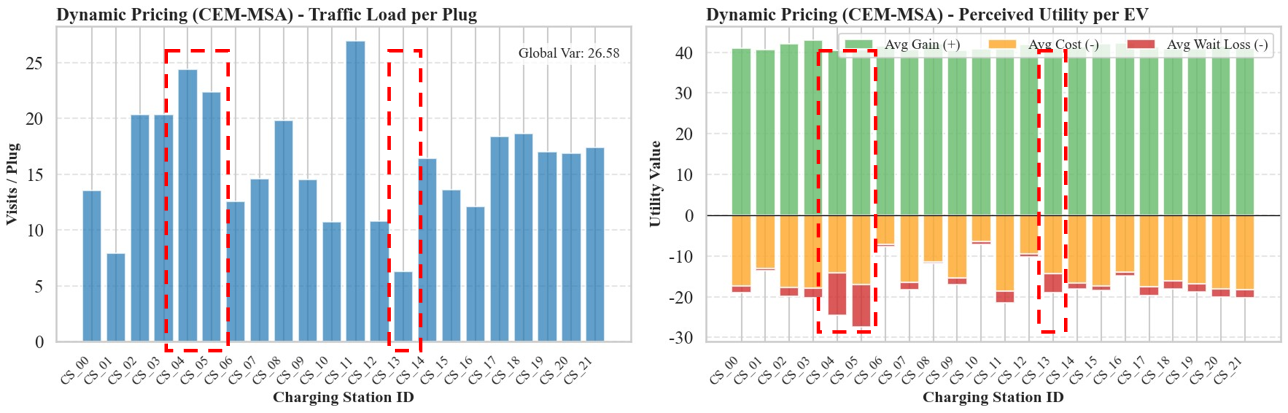}
    \caption{System performance under Dynamic Pricing (CEM-MSA): (Left) Traffic load distribution per plug across charging stations; (Right) Breakdown of perceived utility per EV, including average gain, electricity cost, and waiting loss.}
    \label{fig:traffic_utility_analysis}
\end{figure}

Different charger types imply different service rates for EV users. In the experimental setting, two types of chargers are considered: Fast Charging Stations (FCSs) and Level~2 chargers. FCSs provide higher charging speeds, whereas Level~2 chargers operate at a lower service rate. The comparison between these two charger types under the proposed CEM-MSA framework is illustrated in Figure~\ref{fig:traffic_utility_analysis}.

The left panel of Figure~\ref{fig:traffic_utility_analysis} shows the equilibrium traffic load (i.e., EV charging demand) across all charging stations, where the stations highlighted by the red boxes correspond to Level~2 chargers, while the remaining stations are FCSs. From the traffic loading perspective, existing charging demand dominates congestion patterns. As a result, CS~4, CS~5, and CS~11 exhibit the highest traffic loads, whereas CS~1 and CS~13 show the lowest demand among all stations.

In contrast, the right panel reveals a pronounced discrepancy when examining EV perceived utility compared with the traffic loading pattern observed on the left. The perceived utility of EV users consists of three components: (i) the positive utility gained from completing the charging process, (ii) the negative utility associated with electricity purchasing costs, and (iii) the negative utility caused by the average waiting time at charging stations. A particularly significant difference is observed in the average waiting loss component. Specifically, excluding CS~4, CS~5, and CS~13, most charging stations exhibit relatively small waiting losses, generally ranging from $-1$ to $-3$. However, for the three Level~2 charging stations, the waiting loss increases substantially, reaching values between approximately $-5$ and $-9$.

These results indicate that, although dynamic pricing can partially regulate charging demand, the inherent service rate limitations of lower-speed chargers inevitably lead to longer waiting times and higher disutility for EV users. Although the high waiting losses observed at CS~4 and CS~5 can be partially attributed to their high charging demand, a comparison with another high-demand station such as CS~11 suggests that charger type plays a more dominant role. This effect is further highlighted by CS~13, which has one of the lowest traffic loads among all stations but still experiences the third-highest waiting loss. Overall, charger type heterogeneity primarily manifests through waiting-time-related disutility rather than traffic load alone, underscoring the importance of service rate considerations in pricing-based congestion management.

\subsubsection{Capacity Heterogeneity}

We next examine capacity heterogeneity, which primarily affects system performance by altering the scale of simultaneous charging demand that a station can accommodate, rather than service rates. In this analysis, the five stations with the largest capacities (Stations~6, 8, 10, 12, and~13) are selected and compared against the average dynamic pricing strategy of all remaining stations. This selection is motivated by the PSA-based sensitivity analysis embedded in the PSA-CEM procedure, where factors with higher pricing sensitivity are prioritized due to their stronger influence on pricing outcomes.

\begin{figure}[htbp]
    \centering
    \includegraphics[width=1.0\linewidth]{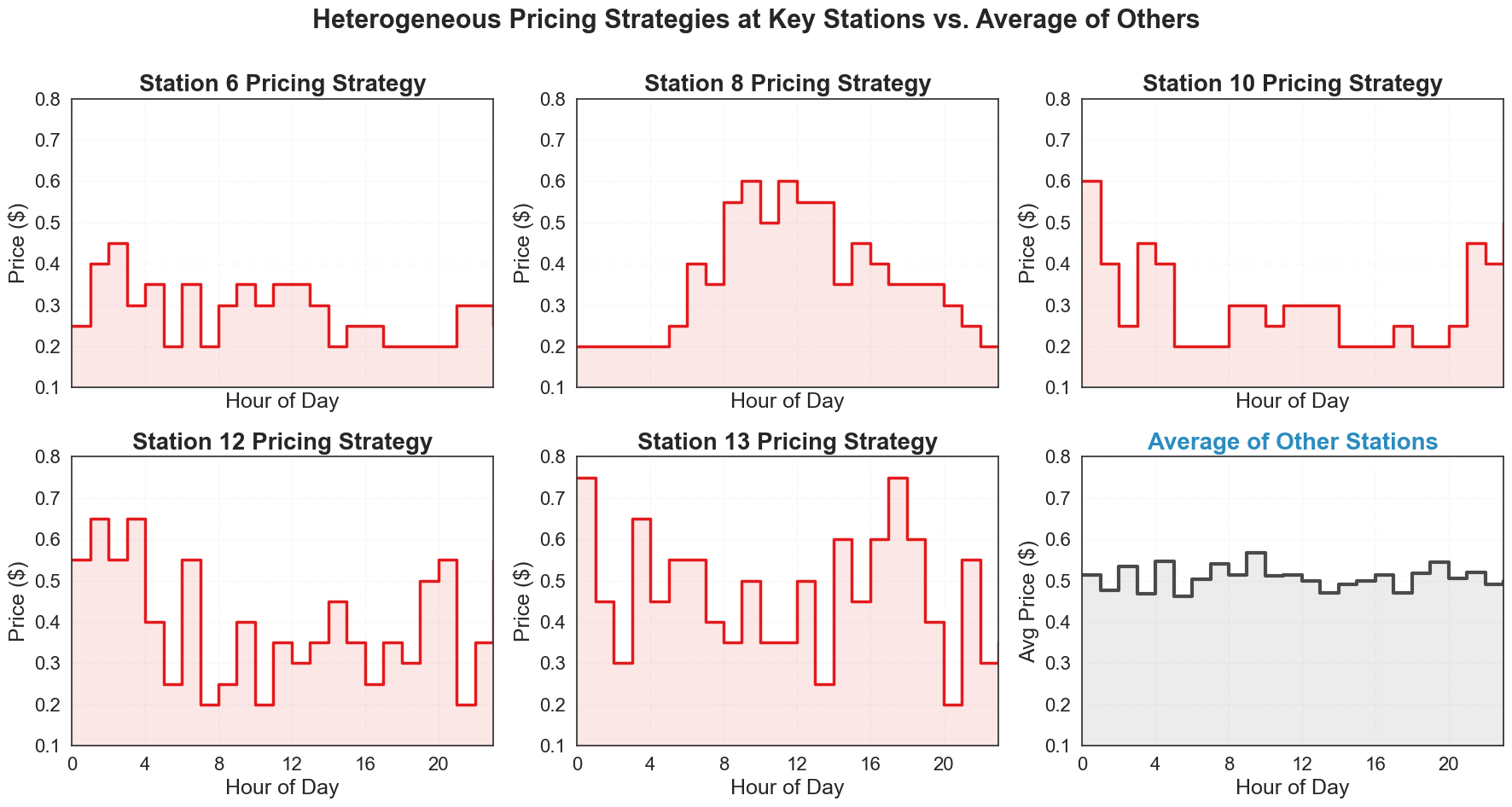}
    \caption{Heterogeneous pricing strategies for selected high-capacity stations (Stations~6, 8, 10, 12, and~13) compared with the average price of all other stations over a 24-hour period.}
    \label{fig:heterogeneous_pricing}
\end{figure}

Figure~\ref{fig:heterogeneous_pricing} illustrates the algorithm’s capability to spatially and temporally differentiate charging prices according to localized demand conditions and station-specific characteristics. Among the selected stations, Station~8 has the highest capacity, with 34 charging plugs, and is located within the Monash University campus. Due to strong daytime charging demand, its pricing strategy exhibits a pronounced increase during daytime hours as a mechanism to counterbalance peak demand and prevent excessive congestion. In contrast, Station~10, which is geographically close to Station~8 but has a much smaller capacity (12 plugs), experiences a spillover or siphoning effect from Station~8. Consequently, its daytime charging price is deliberately reduced to attract EV users and share part of the charging demand, thereby alleviating pressure on Station~8.

Station~6, located at the Chadstone Shopping Centre, adopts a relatively conservative pricing strategy compared with other high-capacity stations. Its charging price remains close to the lower bound (approximately \$0.2) for most of the day, with intermittent price increases occurring between late night (after around 22:00) and midday. This pricing adjustment helps coordinate charging availability with surrounding stations, whose prices generally remain around \$0.5. By strategically adjusting its own price, Station~6 regulates its relative attractiveness to EV users while maintaining lower prices during peak shopping hours in the afternoon and evening to attract EV-driving consumers visiting the shopping centre.

Finally, Stations~12 and~13 both exhibit noticeable temporal price fluctuations, with nighttime charging prices generally exceeding daytime prices. This pattern reflects the fact that higher-capacity stations tend to take on larger charging traffic loads and therefore require stronger incentives to redistribute demand across time and space to relieve system-wide pressure. Compared with Station~12, Station~13 maintains consistently higher prices, reflecting the interaction with charger type heterogeneity discussed earlier. Due to its relatively slower charging speed, the system intentionally discourages excessive EV inflow to Station~13 to avoid excessive waiting losses. Consequently, Station~13 primarily serves as a complementary option to nearby fast charging stations, absorbing overflow demand only when surrounding FCSs experience severe congestion.

\subsection{Dynamic Pricing under the Stackelberg Model: Fixed, Time-of-Use, and Fully Dynamic Pricing}

Finally, we evaluate the effectiveness of the proposed dynamic pricing strategy under the Stackelberg framework by comparing it with two commonly adopted benchmark mechanisms: fixed pricing (traditional pricing), time-of-use (ToU) pricing (semi-dynamic pricing), and the fully dynamic pricing scheme proposed in this study. These mechanisms differ in their degrees of temporal flexibility and their ability to respond to demand variations. Figure~\ref{fig:pricing_comparison} presents an hourly comparison of system performance under the three pricing schemes, while Figure~\ref{fig:total_metrics} provides an aggregated evaluation based on the main utility components.

\subsection{Dynamic Pricing under the Stackelberg Model: Fixed, Time-of-Use, and Fully Dynamic Pricing}

\begin{figure}[htbp]
  \centering
  \includegraphics[width=0.95\linewidth]{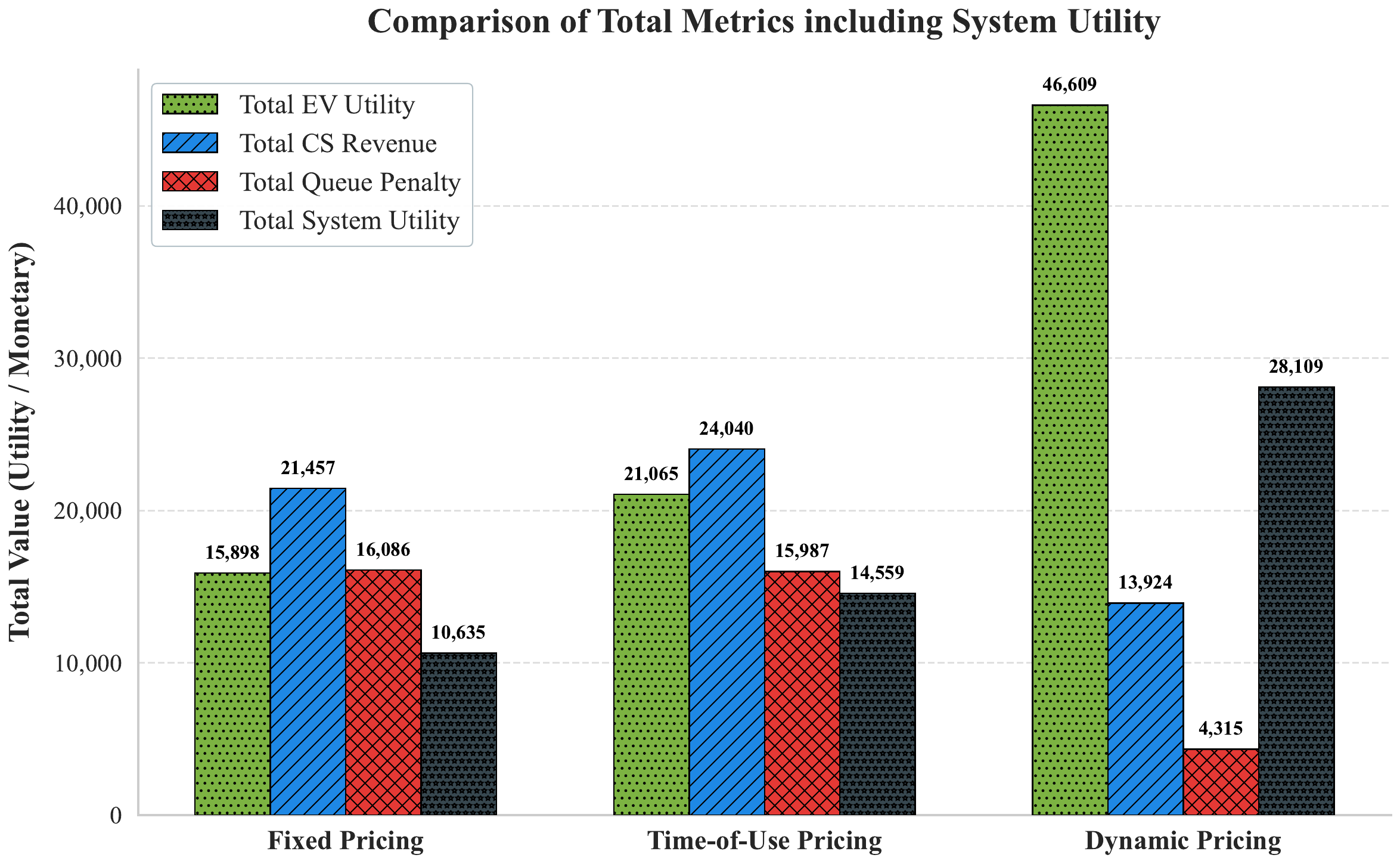}
  
  \caption{Aggregated performance metrics across different pricing strategies.}
  \label{fig:total_metrics}
\end{figure}

Finally, we evaluate the effectiveness of the proposed dynamic pricing strategy under the Stackelberg framework by comparing it with two commonly adopted benchmark mechanisms: fixed pricing (traditional pricing) and time-of-use (ToU) pricing (semi-dynamic pricing). Figure~\ref{fig:pricing_comparison} presents an hourly comparison of system performance, while Figure~\ref{fig:total_metrics} provides an aggregated evaluation based on total utility values.

\subsubsection{Hourly Performance Comparison}

\begin{figure}[htbp]
  \centering
  \includegraphics[width=0.9\linewidth]{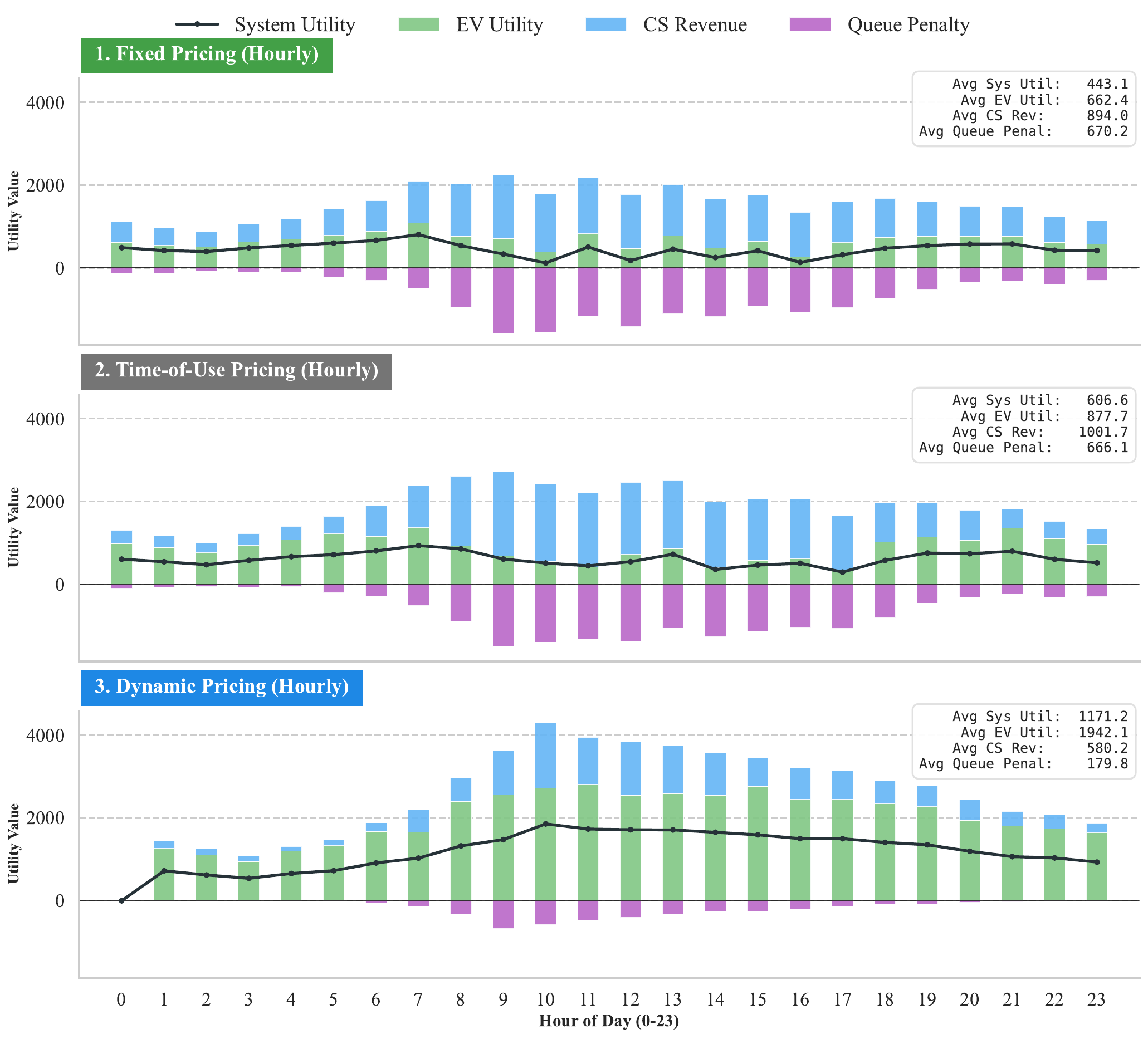}
  
  \caption{Performance comparison of three charging pricing mechanisms. The panels show (1) Fixed Pricing, (2) Time-of-Use Pricing, and (3) Dynamic Pricing over a 24-hour horizon.}
  \label{fig:pricing_comparison}
\end{figure}

Figure~\ref{fig:pricing_comparison} illustrates the temporal evolution of system utility and its components over a 24-hour horizon. Under fixed pricing, charging prices remain constant, resulting in stable CS revenue during off-peak periods but failing to curb demand during peaks. This leads to substantial congestion, indicated by a high average queue penalty of $670.2$ and suboptimal EV utility.

The ToU pricing mechanism partially alleviates this by differentiating between peak and off-peak rates. While it improves average EV utility to $877.7$ and CS revenue to $1001.7$, its coarse temporal structure limits its ability to respond to real-time congestion. Consequently, the average queue penalty remains high at $666.1$, showing little improvement over fixed pricing.

In contrast, the proposed fully dynamic pricing mechanism demonstrates superior adaptability. Although average CS revenue ($580.2$) decreases due to price adjustments, the mechanism drastically reduces congestion. As shown in the top-right legend of Figure~\ref{fig:pricing_comparison}, the average queue penalty drops to $179.8$-a reduction of approximately $73\%$ compared to the benchmarks. This ensures that system utility continues to rise even during high-demand periods, avoiding the performance degradation observed in static approaches.

\subsubsection{Aggregated Utility Comparison and Policy Insights}

Figure~\ref{fig:total_metrics} provides a comprehensive comparison of the accumulated metrics, revealing a distinct trade-off between operator revenue and user welfare. 

While Fixed and ToU pricing strategies secure higher Total CS Revenues ($21,457$ and $24,040$, respectively), they incur severe congestion costs, with Total Queue Penalties reaching approximately $16,000$ in both cases. As a result, the Total System Utility for these mechanisms remains limited at $10,635$ (Fixed) and $14,559$ (ToU).

The proposed dynamic pricing strategy fundamentally shifts this balance. By effectively internalizing congestion externalities, it achieves a massive increase in Total EV Utility-reaching $46,609$, which is more than double the utility provided by ToU pricing ($21,065$). Furthermore, the dynamic approach suppresses Total Queue Penalties to just $4,315$. These improvements culminate in a Total System Utility of $28,109$, representing a $93\%$ increase over ToU pricing and a $164\%$ increase over Fixed pricing.

From a policy perspective, these results highlight that maximizing system welfare does not require prioritizing operator revenue. Instead, redistributing surplus toward EV users through congestion-aware pricing yields substantial system-wide gains. Unlike traditional mechanisms that protect operator profits at the expense of user experience, the proposed Stackelberg-based dynamic pricing creates a balanced outcome by optimizing the trade-off between charging efficiency and congestion management.

\section{Conclusion}

This study proposes a stochastic and heterogeneous dynamic pricing framework for EV charging networks, formulated as a bi-level Stackelberg game in which the charging system operator acts as the leader and EV users respond optimally to pricing signals. By integrating perception errors into the MNL model to represent stochastic EV charging choice behavior and adopting a queuing-theoretic approximation to capture congestion effects, the proposed framework explicitly accounts for charging station heterogeneity and EV user choice stochasticity without relying on deterministic equilibrium assumptions.

The main contributions of this paper are threefold. First, a unified Stackelberg modeling framework is developed to jointly capture dynamic pricing decisions and user choice equilibrium behavior under uncertainty. Compared with fixed and time-of-use pricing schemes, the proposed fully dynamic pricing mechanism effectively internalizes congestion externalities, leading to substantial reductions in queue-related losses and significant improvements in overall system utility.

Second, a rolling-horizon optimization algorithm combining the PSA-CEM with the Method of Successive Averages (MSA) is proposed to efficiently solve the resulting large-scale and time-varying pricing problem. This solution approach preserves the interactive leader and follower structure of the Stackelberg game while ensuring stable convergence and scalability in realistic charging network settings.

Third, through explicit heterogeneity modeling, the study reveals how station-level characteristics and user behavior jointly shape pricing outcomes and system performance. The results demonstrate that charger type heterogeneity primarily affects waiting-time-related disutility through differences in service rates, while capacity heterogeneity drives the spatial and temporal redistribution of charging demand and pricing dynamics. Moreover, the analysis highlights that maximizing charging station revenue does not necessarily maximize system-wide welfare, and that congestion-aware pricing strategies that redistribute surplus toward EV users can yield superior overall performance.

Simulation experiments based on a real-world case study in Clayton, Melbourne, further validate the effectiveness of the proposed framework. The results show that the dynamic pricing mechanism consistently outperforms conventional pricing schemes by reducing congestion, improving EV user utility, and achieving higher total system utility, while exhibiting adaptive responses to heterogeneous infrastructure characteristics and demand patterns.

Despite these contributions, several limitations remain. The current analysis assumes that demand patterns within each rolling horizon are known, whereas real-world EV arrivals may exhibit higher levels of uncertainty and information incompleteness. In addition, heterogeneity is primarily represented through EV user choice behavior and charging station characteristics, while other forms of behavioral heterogeneity, such as heterogeneous driving patterns, are not explicitly embedded in the pricing decision process. Although such heterogeneity may significantly influence optimal pricing outcomes, its impact is difficult to quantify a priori and is therefore examined mainly through ex-post analysis. Furthermore, the study focuses on a single-operator setting and does not consider competitive interactions among multiple charging service providers.

Future research can extend this work in several directions. Incorporating stochastic demand learning and real-time information updates would enable more robust and adaptive pricing strategies. Extending the framework to multi-operator or competitive market environments would allow for the analysis of strategic interactions among charging networks. Additional sources of heterogeneity, such as vehicle battery characteristics, user income levels, charging flexibility, and heterogeneous driving behaviors, could further enrich behavioral realism. Through systematic experimental comparisons across different heterogeneity settings, the effects of heterogeneity could be more explicitly quantified. Finally, integrating power grid constraints, renewable energy availability, and vehicle-to-grid interactions represents a promising direction for designing sustainable and grid-aware EV charging systems.

\bibliographystyle{elsarticle-harv}
\bibliography{references}

@article{Kazemtarghi2024,
  author  = {Kazemtarghi, Amirhossein and Mallik, Avijit and Chen, Yushi},
  title   = {Dynamic pricing strategy for electric vehicle charging stations to distribute the congestion and maximize the revenue},
  journal = {International Journal of Electrical Power \& Energy Systems},
  volume  = {158},
  pages   = {109946},
  year    = {2024}
}

@article{Lee2019,
  author  = {Lee, Wenchao and Schober, Robert and Wong, Vincent W. S.},
  title   = {An analysis of price competition in heterogeneous electric vehicle charging stations},
  journal = {IEEE Transactions on Smart Grid},
  volume  = {10},
  number  = {4},
  pages   = {3990--4002},
  year    = {2019}
}

@article{Lu2022,
  author  = {Lu, Yan and Liang, Ying-Chang and Ding, Zhuo-Hua and Wu, Qiuwei and Ding, Tao and Lee, Wei-Jen},
  title   = {Deep reinforcement learning-based charging pricing for autonomous mobility-on-demand system},
  journal = {IEEE Transactions on Smart Grid},
  volume  = {13},
  number  = {2},
  pages   = {1412--1426},
  year    = {2022}
}

@article{Mi2023,
  author  = {Mi, Hongxuan and Chen, Siyang and Ping, Jinyu and Yan, Zheng},
  title   = {Traffic equilibrium considering heterogeneity across electric vehicles},
  journal = {IEEE Transactions on Intelligent Transportation Systems},
  volume  = {24},
  number  = {12},
  pages   = {14944--14956},
  year    = {2023}
}

@article{Yu2021,
  author  = {Yu, Yue and Su, Chun and Tang, Xiaojun and Kim, Byung-Seo and Song, Tian and Han, Zhu},
  title   = {Hierarchical game for networked electric vehicle public charging under time-based billing model},
  journal = {IEEE Transactions on Intelligent Transportation Systems},
  volume  = {22},
  number  = {1},
  pages   = {518--530},
  year    = {2021}
}

@article{Zhou2021,
  author  = {Zhou, Zhaohui and Moura, Scott J. and Zhang, Hongcai and Zhang, Xinyi and Guo, Qinglai and Sun, Hongbin},
  title   = {Power-traffic network equilibrium incorporating behavioral theory: A potential game perspective},
  journal = {Applied Energy},
  volume  = {289},
  pages   = {116703},
  year    = {2021}
}

@inproceedings{BenGharbia2023,
  author    = {Ben Gharbia, Ibtissem and De Nunzio, Giovanni and Sciarretta, Antonio},
  title     = {Optimal placement of fast charging infrastructure for electric vehicles: An optimal routing and spatial clustering approach},
  booktitle = {Proceedings of the IEEE International Conference on Intelligent Transportation Systems (ITSC)},
  pages     = {2685--2692},
  year      = {2023},
  address   = {Bilbao, Spain}
}

@inproceedings{Patel2022,
  author    = {Patel, Pranav and Kakani, Hitesh and Chaturvedi, Mayank and Kumar, Neeraj},
  title     = {On pricing models for electric vehicle charging},
  booktitle = {Proceedings of the IEEE International Conference on Intelligent Transportation Systems (ITSC)},
  pages     = {158--163},
  year      = {2022},
  address   = {Macau, China}
}

@inproceedings{Amilia2022,
  author    = {Amilia, Nur and Palinrungi, Zulkifli and Vanany, Iwan and Arief, Muhammad},
  title     = {Designing an optimized electric vehicle charging station infrastructure for urban area: A case study from Indonesia},
  booktitle = {Proceedings of the IEEE International Conference on Intelligent Transportation Systems (ITSC)},
  pages     = {2812--2817},
  year      = {2022},
  address   = {Macau, China}
}

@article{Shafiei2023,
  author  = {Shafiei, Mahdi and Ghasemi-Marzbali, Ali},
  title   = {Electric vehicle fast charging station design by considering probabilistic model of renewable energy source and demand response},
  journal = {Energy},
  volume  = {267},
  pages   = {126545},
  year    = {2023}
}

@article{Tran2021a,
  author  = {Tran, Chinh Quy and Keyvan-Ekbatani, Mahdi and Ngoduy, Dong and Watling, David},
  title   = {Stochasticity and environmental cost inclusion for electric vehicles fast-charging facility deployment},
  journal = {Transportation Research Part E: Logistics and Transportation Review},
  volume  = {154},
  pages   = {102460},
  year    = {2021}
}

@inproceedings{Ali2023,
  author    = {Ali, Muhammad Soban and Tangirala, Naga Teja and Knoll, Alois and Eckhoff, David},
  title     = {Rebalancing autonomous electric vehicles for mobility-on-demand by data-driven model predictive control},
  booktitle = {Proceedings of the IEEE International Conference on Intelligent Transportation Systems (ITSC)},
  pages     = {215--221},
  year      = {2023},
  address   = {Bilbao, Spain}
}

@article{Tran2021b,
  author  = {Tran, Chinh Quy and Ngoduy, Dong and Keyvan-Ekbatani, Mahdi and Watling, David},
  title   = {A user equilibrium-based fast-charging location model considering heterogeneous vehicles in urban networks},
  journal = {Transportmetrica A: Transport Science},
  volume  = {17},
  number  = {4},
  pages   = {439--461},
  year    = {2021}
}

@article{Tran2023,
  author  = {Tran, Chinh Quy and Keyvan-Ekbatani, Mahdi and Ngoduy, Dong},
  title   = {Towards clean transportation systems: Infrastructure planning for EVs charging while driving},
  journal = {Sustainable Cities and Society},
  volume  = {96},
  pages   = {104633},
  year    = {2023}
}

@inproceedings{Islam2016,
  author    = {Islam, Md. Shafiul and Nadarajah, Mithulananthan},
  title     = {Daily EV load profile of an EV charging station at business premises},
  booktitle = {Proceedings of the IEEE International Conference on Innovative Smart Grid Technologies - Asia (ISGT-Asia)},
  pages     = {787--792},
  year      = {2016}
}

@article{Jeroslow1985,
  author  = {Jeroslow, Robert G.},
  title   = {The polynomial hierarchy and a simple model for competitive analysis},
  journal = {Mathematical Programming},
  volume  = {32},
  number  = {2},
  pages   = {146--164},
  year    = {1985}
}

@book{rubinstein2004crossentropy,
  author    = {Rubinstein, Reuven Y. and Kroese, Dirk P.},
  title     = {The Cross-Entropy Method: A Unified Approach to Combinatorial Optimization, Monte-Carlo Simulation and Machine Learning},
  publisher = {Springer},
  year      = {2004}
}

@inproceedings{grigorev2021evimpact,
  author    = {Grigorev, Artur and Mao, Tuo and Berry, Adam and Tan, Jency and Purushothaman, Lakshmitha and Mihaita, Adriana-Simona},
  title     = {How will electric vehicles affect traffic congestion and energy consumption: an integrated modelling approach},
  booktitle = {Proceedings of the IEEE International Conference on Intelligent Transportation Systems (ITSC)},
  pages     = {1635--1642},
  year      = {2021},
  address   = {Indianapolis, IN, USA}
}

@misc{IEA2025GlobalEV,
  author       = {{International Energy Agency}},
  title        = {Global EV Outlook 2025},
  year         = {2025},
  howpublished = {\url{https://www.iea.org/reports/global-ev-outlook-2025}},
  note         = {IEA, Paris. Licence: CC BY 4.0}
}

@misc{afirev2022observatoire,
  author       = {{Afirev}},
  title        = {Observatoire de la qualité de service de recharge électrique accessibles au public},
  year         = {2022},
  howpublished = {Online},
  note         = {Accessed: Oct. 05, 2022},
  url          = {https://www.observatoirerecharge-afirev.fr/}
}

@article{feng2023evfcsplanning,
  author  = {Feng, Jiao and Hu, Zechun and Duan, Xiaoyu},
  title   = {EV Fast Charging Station Planning Considering Competition Based on Stochastic Dynamic Equilibrium},
  journal = {IEEE Transactions on Industry Applications},
  volume  = {59},
  number  = {3},
  pages   = {3795--3809},
  year    = {2023}
}

@article{cui2023drlpricing,
  author  = {Cui, Li and Wang, Qingyuan and Qu, Hongquan and Wang, Mingshen and Wu, Yile and Ge, Le},
  title   = {Dynamic Pricing for Fast Charging Stations with Deep Reinforcement Learning},
  journal = {Applied Energy},
  volume  = {346},
  pages   = {121334},
  year    = {2023}
}

@article{Wang2016,
  author  = {Wang, Qiang and Liu, Xiaomei and Du, Jun and Kong, Fanmin},
  title   = {Smart Charging for Electric Vehicles: A Survey From the Algorithmic Perspective},
  journal = {IEEE Communications Surveys \& Tutorials},
  volume  = {18},
  number  = {2},
  pages   = {1500--1517},
  year    = {2016}
}

@article{Zhang2024,
  author  = {Zhang, Yongqi and Fu, Xiao and Yu, Zhuo and Luo, Siyu},
  title   = {How does multi-modal travel enhance tourist attraction accessibility? A refined two-step floating catchment area method using multi-source data},
  journal = {Transactions in GIS},
  volume  = {28},
  pages   = {278--302},
  year    = {2024}
}

@article{Chang2024,
  author  = {Chang, Wencheng and Zhang, Yongqi and Fu, Xiao},
  title   = {Assessment of green space accessibility incorporating sentiment analysis: An improved 2SFCA method},
  journal = {Journal of Geo-Information Science},
  volume  = {26},
  number  = {10},
  pages   = {2243--2253},
  year    = {2024}
}

@article{Bayram2015,
  author  = {Bayram, I. Safak and Tajer, Ali and Abdallah, Mohamed and Qaraqe, Khalid},
  title   = {Capacity Planning Frameworks for Electric Vehicle Charging Stations With Multiclass Customers},
  journal = {IEEE Transactions on Smart Grid},
  volume  = {6},
  number  = {4},
  pages   = {1934--1943},
  year    = {2015}
}

@article{Saharan2020,
  author  = {Saharan, Sanchit and Bawa, Seema and Kumar, Neeraj},
  title   = {Dynamic pricing techniques for Intelligent Transportation System in smart cities: A systematic review},
  journal = {Computer Communications},
  volume  = {150},
  pages   = {603--625},
  year    = {2020}
}

@article{ngoduy2013,
  author  = {Ngoduy, Dong and Maher, Mike},
  title   = {Calibration of second order traffic models using continuous cross entropy method},
  journal = {Transportation Research Part C: Emerging Technologies},
  volume  = {24},
  pages   = {102--121},
  year    = {2013}
}

@article{Zhong2016,
  author  = {Zhong, Renxin and Fu, K. Y. and Sumalee, Agachai and Ngoduy, Dong and Lam, William H. K.},
  title   = {A cross-entropy method and probabilistic sensitivity analysis framework for calibrating microscopic traffic models},
  journal = {Transportation Research Part C: Emerging Technologies},
  volume  = {63},
  pages   = {147--169},
  year    = {2016}
}

@article{Gupta2023,
  author  = {Gupta, Ankit and Bhatnagar, Manav},
  title   = {A Comprehensive Pricing-Based Scheme for Charging of Electric Vehicles},
  journal = {IEEE Systems Journal},
  volume  = {17},
  number  = {3},
  pages   = {4073--4084},
  year    = {2023}
}

@article{Song2025,
  author  = {Song, Yifan and Ngoduy, Dong and Dantsuji, T. and Ding, C.},
  title   = {Dynamic Equilibrium of the Coupled Transportation and Power Networks Considering Electric Vehicles Charging Behavior},
  journal = {Transportation Research Part A: Policy and Practice},
  volume  = {199},
  pages   = {104590},
  year    = {2025}
}

@inproceedings{Qin2011,
  author    = {Qin, Hua and Zhang, Wensheng},
  title     = {Charging scheduling with minimal waiting in a network of electric vehicles and charging stations},
  booktitle = {Proceedings of the 8th ACM International Workshop on Vehicular Inter-Networking (VANET)},
  pages     = {51--60},
  year      = {2011}
}

@article{Wang2021,
  author  = {Wang, Yixiang and Bi, Jun and Guan, Wei and Lu, Chengri and Xie, Dongfan},
  title   = {Optimal charging strategy for intercity travels of battery electric vehicles},
  journal = {Transportation Research Part D: Transport and Environment},
  volume  = {96},
  pages   = {102870},
  year    = {2021}
}

@article{Amjad2018,
  author  = {Amjad, Muhammad and Ahmad, Awais and Rehmani, Mubashir Husain and Umer, Tariq},
  title   = {A review of EVs charging: From the perspective of energy optimization, optimization approaches, and charging techniques},
  journal = {Transportation Research Part D: Transport and Environment},
  volume  = {62},
  pages   = {386--417},
  year    = {2018}
}

@article{Popiolek2023,
  author  = {Popiolek, Alexandre and Dimitrova, Zlatina and Hassler, Jean and Petit, Marc and Dessante, Philippe},
  title   = {Comparison of decentralised fast-charging strategies for long-distance trips with electric vehicles},
  journal = {Transportation Research Part D: Transport and Environment},
  volume  = {124},
  pages   = {103953},
  year    = {2023}
}

@article{Sheng2021,
  author  = {Sheng, Yujie and Guo, Qinglai and Chen, Fang and Xu, Lixiong and Zhang, Yi},
  title   = {Coordinated pricing of coupled urban power-traffic networks: The value of information sharing},
  journal = {Applied Energy},
  volume  = {301},
  pages   = {117428},
  year    = {2021}
}

@article{Hu2021,
  author  = {Hu, Xiaowei and Yang, Zhongzhen and Sun, Jian and Zhang, Yaling},
  title   = {Sharing economy of electric vehicle private charge posts},
  journal = {Transportation Research Part B: Methodological},
  volume  = {152},
  pages   = {258--275},
  year    = {2021}
}

@article{Hu2023,
  author  = {Hu, Xiaowei and Yang, Zhongzhen and Sun, Jian and Zhang, Yaling},
  title   = {Optimal pricing strategy for electric vehicle battery swapping: Pay-per-swap or subscription?},
  journal = {Transportation Research Part E: Logistics and Transportation Review},
  volume  = {171},
  pages   = {103030},
  year    = {2023}
}

@article{Wang2025,
  author  = {Wang, Yanyan and Fan, Ruoguang and Lin, Jioucha and Xie, Xiaoxuan and Zhang, Weijian and Srinivasan, Dipti},
  title   = {Subsidies for shared private electric vehicle chargers: A three-level Stackelberg game analysis},
  journal = {Transportation Research Part D: Transport and Environment},
  volume  = {140},
  pages   = {104626},
  year    = {2025}
}

@article{Zheng2019,
  author  = {Zheng, Yelin and Song, Yan and Hill, David J. and Meng, Ke},
  title   = {Online distributed optimal scheduling for EV charging stations in distribution systems based on distributed MPC},
  journal = {IEEE Transactions on Industrial Informatics},
  volume  = {15},
  number  = {2},
  pages   = {638--649},
  year    = {2019}
}

@article{Yin2024,
  author  = {Yin, Linfei and Zhang, Yunfei},
  title   = {Particle swarm optimization based on data driven for EV charging station siting},
  journal = {Energy},
  volume  = {310},
  pages   = {133197},
  year    = {2024}
}

@article{Park2024,
  author  = {Park, Hyunwoo and Lee, Chungmok},
  title   = {An exact algorithm for maximum electric vehicle flow coverage problem with heterogeneous chargers, nonlinear charging time and route deviations},
  journal = {European Journal of Operational Research},
  volume  = {315},
  number  = {3},
  pages   = {926--951},
  year    = {2024}
}

@article{Calik2021,
  author  = {Çalık, Hatice and Oulamara, Ammar and Prodhon, Caroline and Salhi, Said},
  title   = {The electric location-routing problem with heterogeneous fleet: Formulation and Benders decomposition approach},
  journal = {Computers \& Operations Research},
  volume  = {131},
  pages   = {105251},
  year    = {2021}
}

@article{Zheng2020,
  author  = {Zheng, Nan and Geroliminis, Nikolas},
  title   = {Area-based equitable pricing strategies for multimodal urban networks with heterogeneous users},
  journal = {Transportation Research Part A: Policy and Practice},
  volume  = {136},
  pages   = {357--374},
  year    = {2020}
}

@article{Mao2024,
  author  = {Mao, Shiyue and Jin, Jishuo and Xu, Yinliang},
  title   = {Routing and Charging Scheduling for EV Battery Swapping Systems: Hypergraph-Based Heterogeneous Multiagent Deep Reinforcement Learning},
  journal = {IEEE Transactions on Smart Grid},
  volume  = {15},
  number  = {5},
  pages   = {4903--4916},
  year    = {2024}
}

@inproceedings{Zhang2022,
  author    = {Zhang, Weijia and Liu, Hao and Han, Jindong and Ge, Yong and Xiong, Hui},
  title     = {Multi-Agent Graph Convolutional Reinforcement Learning for Dynamic Electric Vehicle Charging Pricing},
  booktitle = {Proceedings of the 28th ACM SIGKDD Conference on Knowledge Discovery and Data Mining},
  pages     = {2471--2481},
  year      = {2022},
  address   = {Washington DC, USA}
}

@article{Mrkos2022,
  author  = {Mrkos, Jan and Basmadjian, Robert},
  title   = {Dynamic Pricing for Charging of EVs with Monte Carlo Tree Search},
  journal = {Smart Cities},
  volume  = {5},
  number  = {1},
  pages   = {223--240},
  year    = {2022}
}

@article{Elkholy2026,
  author  = {Elkholy, Ahmed M. and Rozhkov, Alexander N. and Badalyan, Artush V. and Cherdintsev, Ivan A.},
  title   = {Adaptive Genetic Algorithms Enhance EV Charging Infrastructure Resilience through Multi-Constraint Optimization of Grid Resources and Traffic Dynamics},
  journal = {Electric Power Systems Research},
  volume  = {250},
  pages   = {112045},
  year    = {2026}
}

@article{Kullman2021,
  author  = {Kullman, Nicholas D. and Goodson, Justin C. and Mendoza, Jorge E.},
  title   = {Electric Vehicle Routing with Public Charging Stations},
  journal = {Transportation Science},
  volume  = {55},
  number  = {3},
  pages   = {637--659},
  year    = {2021}
}

@article{Liu2006,
  author  = {Liu, Hao and Chen, Wei and Sudjianto, Agus},
  title   = {Relative Entropy Based Method for Probabilistic Sensitivity Analysis in Engineering Design},
  journal = {Journal of Mechanical Design},
  volume  = {128},
  number  = {2},
  pages   = {326--336},
  year    = {2006}
}

@article{Luo2018Stochastic,
  author  = {Luo, Chao and Huang, Yih-Fang and Gupta, Vijay},
  title   = {Stochastic Dynamic Pricing for {EV} Charging Stations With Renewable Integration and Energy Storage},
  journal = {IEEE Transactions on Smart Grid},
  volume  = {9},
  number  = {2},
  pages   = {1494--1505},
  year    = {2018}
}

@article{Soares2017Dynamic,
  author  = {Soares, Jo{\~a}o and Ghazvini, Mohammad Ali Fotouhi and Borges, Nuno and Vale, Zita},
  title   = {Dynamic electricity pricing for electric vehicles using stochastic programming},
  journal = {Energy},
  volume  = {122},
  pages   = {111--127},
  year    = {2017}
}

@article{Wu2019Stochastic,
  author  = {Wu, Fei and Sioshansi, Ramteen},
  title   = {A stochastic operational model for controlling electric vehicle charging to provide frequency regulation},
  journal = {Transportation Research Part D: Transport and Environment},
  volume  = {67},
  pages   = {475--490},
  year    = {2019}
}

@article{Zanvettor2022Stochastic,
  author  = {Zanvettor, Giovanni G. and Casini, Marco and Smith, Roy S. and Vicino, Antonio},
  title   = {Stochastic Energy Pricing of an Electric Vehicle Parking Lot},
  journal = {IEEE Transactions on Smart Grid},
  volume  = {13},
  number  = {4},
  pages   = {3069--3081},
  year    = {2022}
}

@book{Saltelli2008,
  author    = {Saltelli, Andrea and Ratto, Marco and Andres, Terry and Campolongo, Francesca and Cariboni, Jessica and Gatelli, Debora and Saisana, Michaela and Tarantola, Stefano},
  title     = {Global Sensitivity Analysis: The Primer},
  publisher = {John Wiley \& Sons},
  year      = {2008}
}

@article{Chen2024Etiquette,
  author  = {Chen, Bingkun and Chen, Zhuo and Liu, Xiaoyue Cathy and Yi, Zhiyan},
  title   = {Bayesian inference-based spatiotemporal modeling with interim activities for EV charging etiquette},
  journal = {Transportation Research Part D: Transport and Environment},
  volume  = {127},
  pages   = {104060},
  year    = {2024}
}

@article{WANG2025104666,
  author  = {Wang, Yucheng and Yang, Min and Qin, Bozhan and Zhang, Yongqi},
  title   = {Decoding travel behavioral intentions under flight delays via interpretable machine learning: Insights for safeguarding passenger mobility},
  journal = {Transportation Research Part A: Policy and Practice},
  volume  = {201},
  pages   = {104666},
  year    = {2025}
}

\clearpage  
\appendix  
\section{Sensitivity Analysis of the System Utility Weights}
\label{appendix:sensitivity_analysis}

\begin{figure}[htbp]
  \centering
  \includegraphics[width=0.9\linewidth]{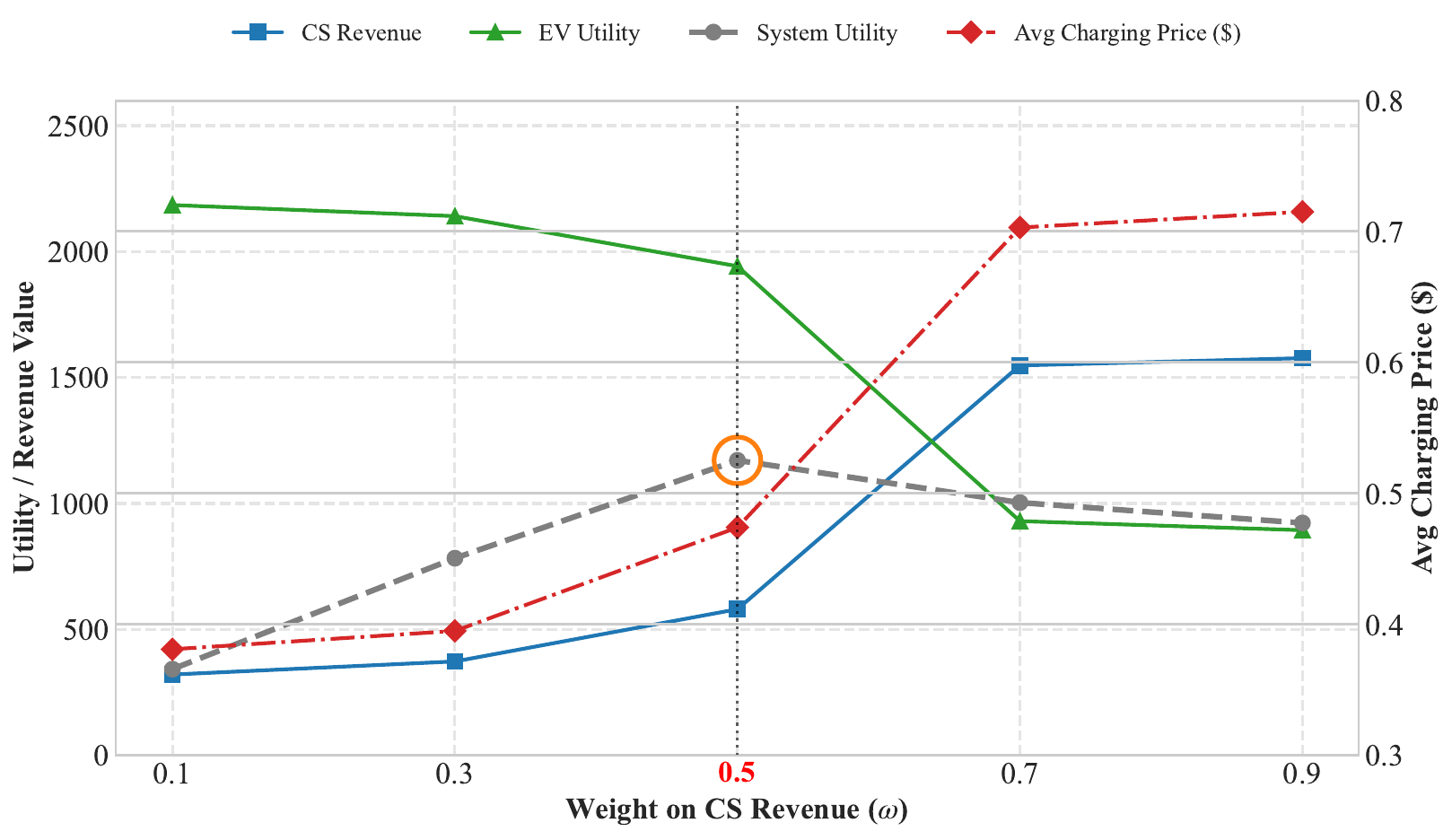}
  \caption{Sensitivity analysis of the weight parameter $\omega$ on system performance metrics. The dual Y-axes display utility/revenue values (left) and average charging prices (right). The orange circle highlights the global maximum system utility achieved at the balanced scenario ($\omega=0.5$).}
  \label{fig:sensitivity_omega}
\end{figure}

In this sensitivity analysis, we conduct five experimental scenarios by varying the system utility weight parameter $\omega \in \{0.1, 0.3, 0.5, 0.7, 0.9\}$, spanning a continuum from an EV user-oriented regime ($\omega=0.1$) to an operator revenue-oriented regime ($\omega=0.9$). This design enables a systematic examination of how the relative emphasis on stakeholder objectives shapes pricing decisions and overall system performance.

As illustrated in Figure~\ref{fig:sensitivity_omega}, variations in $\omega$ induce a pronounced trade-off between EV user utility and CS revenue, mediated through the endogenous pricing mechanism. Specifically, as $\omega$ increases, the optimization process progressively prioritizes operator revenue, resulting in a steady escalation of the average charging price from \$0.38 at $\omega=0.1$ to \$0.72 at $\omega=0.9$. While this pricing strategy effectively enhances CS revenue (blue line), it simultaneously imposes increasing disutility on EV users, leading to a sharp and monotonic decline in EV utility (green line). This opposing response clearly reflects the inherent conflict between affordability for users and profitability for operators in dynamic charging markets.

Importantly, the aggregate System Utility (grey dashed line) does not follow a linear interpolation between these two extremes. Instead, it exhibits a distinct concave pattern, suggesting the presence of an interior optimum. Extreme weight configurations are associated with suboptimal system outcomes. At $\omega=0.1$, excessively low charging prices disproportionately favor EV users, but insufficient operator revenue undermines the sustainability of charging infrastructure provision, resulting in a reduced overall system utility. Conversely, at $\omega=0.9$, aggressive revenue-driven pricing suppresses effective demand and substantially erodes user welfare, again leading to a deterioration in total system performance.

The balanced scenario ($\omega=0.5$) corresponds to the apex of the system utility curve, highlighted by the orange circle. This outcome provides an empirical validation for the chosen weighting scheme, demonstrating that an equal-weight configuration is not merely a heuristic or normative assumption, but rather emerges endogenously as the global optimum within the explored parameter space. At this point, marginal gains in operator revenue are closely aligned with marginal losses in user utility, yielding the most efficient compromise between economic viability and service affordability.

Further deviations from this equilibrium exhibit diminishing returns. Increasing $\omega$ beyond 0.5 leads to revenue improvements that are more than offset by disproportionate losses in EV user utility, resulting in a rapid decline in social welfare. In contrast, decreasing $\omega$ below 0.5 produces revenue shortfalls that threaten operator viability and long-term system resilience. Taken together, these findings indicate that $\omega=0.5$ represents an empirically optimal operating point that reconciles conflicting stakeholder objectives and maximizes overall system efficiency. Therefore, based on the above empirical evidence, $\omega=0.5$ is adopted as the weight coefficient in the subsequent analysis.

\section{Cross-Entropy Method for Optimization}
\label{appendix:cem}

Without loss of generality, consider the minimization of the objective function $F(\boldsymbol{\theta})$ 
over the decision vector $\boldsymbol{\theta}$ within the feasible domain $\Omega$:

\begin{equation}
    F^* = \min_{\boldsymbol{\theta} \in \Omega} F(\boldsymbol{\theta}),
    \label{eq:cem_min}
\end{equation}
where $F^*$ represents the minimum of the objective function, and $\boldsymbol{\theta}$ 
denotes the decision vector within the feasible domain $\Omega$.

The CEM introduces a family of parameterized probability density functions (PDFs) 
$f(\boldsymbol{\theta}; \boldsymbol{\phi})$ to characterize the sampling distribution. 
For the minimization of $F(\boldsymbol{\theta})$, a rare event is defined by
$F(\boldsymbol{\theta}) \le c$, where $c \ge F^*$ is a threshold sufficiently close to the optimum. 
The associated probability is given by
\begin{equation}
    \ell(c) = \mathbb{P}_{\boldsymbol{\phi}}(F(\mathbf{\Theta}) \le c)
    = \mathbb{E}_{\boldsymbol{\phi}}\!\left[\mathbb{I}_{\{F(\mathbf{\Theta}) \le c\}}\right],
    \label{eq:cem_prob}
\end{equation}
where $\mathbb{P}_{\boldsymbol{\phi}}(\cdot)$ and $\mathbb{E}_{\boldsymbol{\phi}}(\cdot)$ denote
the probability and expectation operators, respectively, and $\mathbb{I}(\cdot)$ is the indicator function.

The CEM algorithm iteratively updates the sampling distribution to maximize the likelihood of generating elite samples.  
The procedure is summarized as follows:

\begin{enumerate}
    \item Initialize the parameter vector $\boldsymbol{\phi}^{(0)}$ and set the iteration counter $t = 1$.
    \item At each iteration, draw a population of $N$ random samples
    $\boldsymbol{\theta}_1, \boldsymbol{\theta}_2, \ldots, \boldsymbol{\theta}_N$ 
    from the distribution $f(\boldsymbol{\theta}; \boldsymbol{\phi}^{(t-1)})$, and evaluate the objective function $F(\boldsymbol{\theta}_i)$ for each sample.
    \item Sort the objective values and determine the elite threshold:
    \begin{equation}
        \hat{c}^{(t)} = F_{(\lceil (1-q)N \rceil)},
        \label{eq:cem_threshold}
    \end{equation}
    where $q \in (0,1)$ is the elite ratio.
    \item Update the distribution parameters via maximum-likelihood estimation:
    \begin{equation}
        \boldsymbol{\phi}^{(t,\text{new})} 
        = \arg\max_{\boldsymbol{\phi}} 
        \frac{1}{N} 
        \sum_{i=1}^{N} 
        \mathbb{I}_{\{F(\boldsymbol{\theta}_i)\le \hat{c}^{(t)}\}} 
        \ln f(\boldsymbol{\theta}_i; \boldsymbol{\phi}).
        \label{eq:cem_update}
    \end{equation}
    \item Apply smoothing to avoid premature convergence:
    \begin{equation}
        \boldsymbol{\phi}^{(t)} = \beta \boldsymbol{\phi}^{(t,\text{new})} + (1-\beta)\boldsymbol{\phi}^{(t-1)}, 
    \end{equation}
    where $\beta \in [0,1]$ is the smoothing factor.
    \item Check convergence: if the standard deviation of the sampling distribution falls below a tolerance $\epsilon$, set
    \begin{equation}
        F^* = \hat{c}^{(t)};
        \label{eq:cem_stop}
    \end{equation}
    otherwise, increment $t \leftarrow t+1$ and repeat steps 2--5.
\end{enumerate}

By iteratively updating the sampling distribution in this manner, the CEM approximates the optimal solution 
of the original optimization problem while efficiently exploring the feasible space.

\end{document}